\documentclass[twocolumn]{aastex631}
\usepackage{amsmath,amssymb,amsthm}
\usepackage{comment}
\usepackage{graphicx}
\usepackage{natbib}
\usepackage{xcolor}

\defcitealias{metzger2024}{M24}

\shorttitle{AASTeX v6.3.1 Sample article}
\shortauthors{Chen et al.}
\graphicspath{{./}{figures/}}

\begin{document}

\title{Gravitational Instability and Fragmentation in Collapsar Disks Supports the Formation of Sub-Solar Neutron Stars}

\author{Yi-Xian Chen}
\affiliation{Department of Astrophysical Sciences, Princeton University, Princeton, NJ 08544, USA} 

\author[0000-0002-4670-7509]{Brian D.~Metzger}
\affil{Department of Physics and Columbia Astrophysics Laboratory, Columbia University, New York, NY 10027, USA}
\affil{Center for Computational Astrophysics, Flatiron Institute, 162 5th Ave, New York, NY 10010, USA}

\correspondingauthor{Yi-Xian Chen}
\email{yc9993@princeton.edu}

\begin{abstract}

We perform three-dimensional shearing-box hydrodynamical simulations to explore the outcome of gravitational instability in the outer regions of neutrino-cooled disks such as those formed from the collapse of rotating massive stars (``collapsars''). We employ a physical equation of state, optically-thin neutrino cooling, and assume an electron fraction set by the balance of $e^{\pm}$ pair-capture reactions.  Disks in a marginally stable initial state (Toomre parameter $Q \approx 1$) undergo runaway cooling and fragmentation when the dimensionless cooling timescale obeys $\tau_{\rm cool} \equiv t_{\rm cool}\Omega \lesssim 10$, where $\Omega$ is the orbital frequency; these conditions correspond to accretion rates $\gtrsim M_{\odot}$ s$^{-1}$ on the upper end of those achieved by collapsar progenitor stars. Fragmentation leads to the formation of neutron-rich clumps (electron fraction $Y_{e} \lesssim 0.1$) spanning a range of masses $\sim 0.01-1M_{\odot}$ around the local Jeans value. Most clumps exceed the local Chandrasekar mass $M_{\rm Ch} \propto Y_{e}^{2}$ and hence will continue to collapse to nuclear densities, forming neutron stars (NS) with sub-solar masses otherwise challenging to create through ordinary stellar core-collapse. Even cool disks dominated by $\alpha$ particles ($Y_{e} \simeq 0.5$) can fragment and collapse into neutron-rich clumps capable of forming sub-solar NSs. Although our simulations cannot follow this process directly, if the disk-formed NSs subsequently pair into binaries, the gravitational wave chirps from their rapid mergers are potentially detectable by ground based observatories. The temporal coincidence of such a hierarchical NS merger chain with the collapsar gamma-ray burst and supernova would offer a uniquely spectacular multi-messenger ``symphony''.
\end{abstract}


\section{Introduction} \label{sec:intro}

The cores of a small fraction of massive stars rotate rapidly at the end of their nuclear burning evolution.  Although the origin of this rapid terminal rotation remains under debate (e.g., \citealt{Cantiello2007}), particularly in the face of strong outwards angular momentum transport during earlier stages of stellar evolution (e.g., \citealt{Kissin&Thompson18,Fuller+19}), it is necessary to explain some of the most luminous transients in the universe from stellar collapse, namely long-duration gamma-ray bursts (GRB; e.g., \citealt{Stanek+03}) and their often associated hyper-energetic supernovae (e.g., \citealt{Woosley2006}). 

In particular, a star that is slowly rotating upon collapse and fails to explode promptly will form a slowly spinning black hole (BH; e.g., \citealt{Fuller&Ma19}). Most of the stellar envelope falls directly into the horizon, though a fraction may be ejected in a weak outflow giving rise to a dim electromagnetic transient (e.g., \citealt{Antoni&Quataert23}). 
In contrast, if the star possesses enough ordered angular momentum, then its outer layers will instead form a rotationally supported disk around the BH, feeding it at high rates $\sim 0.1 - 10 M_\odot$ s$^{-1}$ and potentially powering a relativistic GRB jet \citep{Woosley1993,Gottlieb+24} and an energetic supernova (e.g., \citealt{Lopez-Camara+09,Dean&Fernandez24}).  Such events, in which a hyper-accreting stellar mass BH plays the central role, are referred to as ``collapsars'' (e.g., \citealt{MacFadyen&Woosley99}). A similar hyper-accreting disk can form around the central compact object if it is a highly magnetized proto-neutron star instead of a BH (e.g., \citealt{Metzger+18b,Obergaulinger&Aloy01}).

Collapsar disks reach sufficiently high temperatures $kT \gtrsim $ MeV to efficiently cool by emitting neutrinos and anti-neutrinos, primarily those produced by the URCA captures of electrons and positrons on free protons and neutrons, respectively (e.g., \citealt{Beloborodov2003}). For high mass inflow rates, the outer regions of neutrino-cooled collapsar disks are susceptible to gravitational instability \citep{Perna+06,ChenBel2007,Lopez-Camara+09,Taylor+11,Liu2017,Shahamat+21}, typically outside hundreds of gravitational radii, where the \citet{Toomre64} parameter $Q$ drops below unity. 

The long-term evolution of self-gravitating disks depends sensitively on thermodynamics (see \citealt{Kratter&Lodato16} for a review).  When the disk cools through radiation or other processes slowly compared to the local dynamical time, the gravitational instability drives turbulence sufficient to heat the disk and regulate its structure to a $Q \sim 1$ state \citep{Gammie2001}. On the other hand, when the cooling rate is faster than dynamical, the disk fragments into self-gravitating ``clumps'', analogous to the formation of gas giants in protoplanetary disks \citep{Gammie2001,Lodato2004,Kratter+10} and stars in the disks of active galactic nuclei (AGN; \citealt{GoodmanTan2004,chen2023}).  In collapsar disks, clumps produced by gravitational instability may collapse to form neutron stars (NS) \citep{Piro&Pfahl07}. 

\citet[hereafter \citetalias{metzger2024}]{metzger2024} 
recently revisited the conditions for collapsar disks to fragment into NSs and suggested 
several novel observational consequences were this to occur. 
In particular,
\citetalias{metzger2024} emphasized that a minimal condition for a self-bound clump to continue to collapse to higher densities 
is that its mass $M_{\rm clump}$ must exceed the local Chandrasekhar mass \citep{Chandrasekhar1931}
\begin{equation}
    M_{\rm Ch} \approx 1.45M_\odot (Y_e/0.5)^2,
    \label{eq:MCh}
\end{equation}
where $Y_e$ is the local electron fraction.  As a clump which obeys $M_{\rm clump} > M_{\rm Ch}$ loses its thermal pressure support due to runaway neutrino cooling, even electron degeneracy pressure cannot halt its collapse. This process is qualitatively similar to the collapse of a star's degenerate iron core at the end of its nuclear burning evolution. 

Unlike the iron core of a massive star ($Y_{e} \approx 0.4-0.5$), however, collapsar disks can become extremely rich in neutrons ($Y_e \sim 0.1$; e.g., \citealt{Metzger+08b}).  This difference results from the higher temperatures and much greater densities (greater electron degeneracy $\mu/kT \gtrsim 1$) of neutrino-cooled collapsar disks, which drive $Y_{e}$ to lower values than achieved in a massive star's iron core at collapse (\citealt{Beloborodov2003}, see their Fig.~1, or the left panel of our Fig.~\ref{fig:Yeeq-rho-T}).  Because $M_{\rm Ch} \ll M_{\odot}$, collapsar disk-formed NSs can in principle possess much lower masses $\sim 0.01-1 M_{\odot}$ than those born of ordinary supernovae (modern core-collapse simulations find a minimum NS mass $\approx 1.2M_{\odot}$; e.g., \citealt{Muller+25}). Even NS below the usually considered minimum stable mass $M_{\rm min} \approx 0.1M_{\odot}$ (e.g., \citealt{Haensel+02}) may be able to form temporarily in this environment (Sec.~\ref{sec:minmass}).

 If collapsar disks fragment into multiple NS, these bodies may find themselves paired into tight binaries, as a result of fissioning from a single collapsing clump (e.g., \citealt{Nesvorny+10}) or through gas drag pairing of initially unbound NS (e.g., \citealt{Dodici&Tremaine24}). Subsequent coalescence of these binaries through gravitational waves (GW) would then offer a novel source of sub-solar compact binary mergers (\citetalias{metzger2024}), potentially detectable by LIGO/Virgo \citep{Abbott_subsolar_22} and future ground-based observatories.  Any compact objects that ultimately survive such merger events (\citetalias{metzger2024}), will eventually coalesce with the central BH, generating a final GW chirp \citep{Piro&Pfahl07,Lerner2025}.

Here we explore the conditions for the fragmentation of self-gravitating collapsar disks using hydrodynamic simulations in a local shearing-box setup, including a physical equation of state and optically-thin neutrino cooling appropriate to the collapsar environment.  Our local simulations do not follow the global disk evolution, much less its self-consistent formation from the collapsing star (as was the focus of several recent simulation works that neglect self-gravity; e.g., \citealt{Gottlieb+23,Dean&Fernandez24,Issa+25}). Nevertheless, our simulations are the first of their kind and provide insights into the conditions necessary for fragmentation and the mass distribution of the resulting bound bodies, i.e., the ``initial mass function'' of collapsar-formed NS. We confirm that when the neutrino cooling is sufficiently rapid, the disk fragments into clumps spanning a range of masses $\sim 0.01-1M_{\odot}$ and that most such clumps exceed the local Chandrasekhar limit, enabling their continued collapse to NSs of similar masses.

This paper is organized as follows. We introduce the setup of our \texttt{Athena++} simulations in Sec.~\ref{sec:setup}. In Sec.~\ref{sec:results} we present results from both steadily-accreting and fragmenting simulations, focusing on the latter case and the resulting properties of self-gravitating clumps. In Sec.~\ref{sec:discussion} we discuss uncertainties and limitations of our calculations as well as their implications for multi-messenger signals from collapsars.  In Sec.~\ref{sec:conclusion} we summarize our conclusions.

\section{Numerical Setup}
\label{sec:setup}

We employ \texttt{Athena++} \citep{Stone2020} to solve the hydrodynamic equations in a three-dimensional Cartesian shearing box configuration coupled with self-gravity. These read:

\begin{equation}
\frac{\partial \rho}{\partial t}+\nabla \cdot(\rho \boldsymbol{v})=0
\end{equation}

\begin{equation}
\begin{aligned}
& \frac{\partial(\rho \boldsymbol{v})}{\partial t}+{\nabla} \cdot\left(\rho \boldsymbol{v} \boldsymbol{v}\right) + \nabla P=-\rho {\nabla} \Phi_{\rm sg} \\
& -2 \rho \Omega \hat{z} \times \boldsymbol{v}+2 q \rho \Omega^2 x \hat{\boldsymbol{x}}-\rho \Omega^2 z \hat{z}
\end{aligned}
\end{equation}

\begin{equation}
\begin{aligned}
\frac{\partial E}{\partial t} & +\nabla \cdot\left(E+P\right) \boldsymbol{v}=-\rho \boldsymbol{v} \cdot \nabla \Phi_{\rm sg} \\
& +\rho \Omega^2 \boldsymbol{v} \cdot(2 q x \hat{\boldsymbol{x}}-z \hat{\boldsymbol{z}})-  \Lambda,
\end{aligned}
\end{equation}
where $\rho$ is the baryon mass density, $E=U +\rho v^2 / 2$ is the sum of internal and kinetic energy, most other variables take on their usual meanings, and we take $q = -d{\rm ln \Omega}/d{\rm r} = 3/2$ for the shear parameter for a Keplerian disk. The gravitational potential is obtained by solving the Poisson equation,
\begin{equation}
   \nabla^2 \Phi_{\rm sg} = 4\pi G \rho, 
\end{equation}
using fast Fourier transforms \citep{Kim2011}, as applied in \citet{chen2023}. 

To close the equation set, we need a general equation of state linking internal energy with pressure $U(P, \rho)$ as well as an expression for the optically-thin neutrino cooling rate $\Lambda(P, \rho)$.  As described in Appendix \ref{app:EOS}, these are calculated for an ideal plasma containing photons, electron/positron pairs (for arbitrary relativism and degeneracy), free nucleons (neutrons and protons), and $\alpha$ particles, following \citet{ChenBel2007, Liu2017, Lerner2025}. The number density of positrons is determined by the equilibrium between pair creation and annihilation reactions $e^{-} + e^{+} \leftrightarrow \gamma + \gamma,$ which is an excellent approximation at the high temperatures of interest.  For marginally Toomre-stable disks that undergo fragmentation, we shall find typical initial midplane densities and temperatures in the range $\rho_0 \sim 10^{8}-10^{9}$ g cm$^{-3}$ and $T_0 \sim 0.5-1.5$ MeV, corresponding to electrons that are relativistic and mildly degenerate.

An important property of the disk material is the electron fraction,
\begin{equation}
    Y_{e} \equiv \frac{n_{\rm p}}{\rho/m_p}
\end{equation}
where $n_{\rm p}$ is the number density of protons, which includes both free nucleons and protons locked in alpha particles, and $m_p$ is the proton mass. Rather than following the time-evolution of $Y_e$ self-consistently, we assume that it everywhere reaches an equilibrium value $Y_{e, \rm eq}$ (with an associated electron chemical potential $\mu  = \mu_{\rm eq}$) set by the balance of electron captures on protons and positron captures on neutrons \citep{Beloborodov2003}.
This is generally a good approximation because these capture reactions in neutrino-cooled disks are typically fast compared to the gas inflow time in dense regions of the disk where self-gravity is relevant \citep{ChenBel2007}.  As shown in Appendix \ref{app:EOS} (Fig.~\ref{fig:tcool}, right panel), the electron capture timescale (over which $Y_{e}$ will find its equilibrium value) is shorter than the neutrino cooling timescale (over which fragmentation will take place) across most densities and temperatures. We neglect the effects of $\nu_{e}$ and $\bar{\nu}_{e}$ captures by free nucleons on $Y_{e,eq}$ because they are generally slow compared to electron/positron captures in the disk interior.

We assume nuclear statistical equilibrium (NSE) in calculating the $\alpha$ particle mass fraction, $X_\alpha(\rho,T,Y_e)$. At high temperatures all nucleons are free, while at lower temperatures $\alpha$ particles form up to the maximum mass fraction $X_{\alpha} = 2Y_{e, \rm eq}$ allowed by the electron fraction.  These two regimes are roughly separated by the following $\rho-T$ contour (Eq.~\eqref{eqn:GrhoT}):
\begin{equation}
    \mathcal{G}(\rho, T) = 4.9 \times 10^2 \rho_{10}^{-3 / 2} T_{10}^{9 / 4} \exp \left(-\frac{16.4}{T_{10}}\right) = 1, 
    \label{eqn:Gmaintext}
\end{equation}
where $\rho_{10} = \rho/(10^{10}$g/cm$^3)$ and $T_{10} = T/(10^{10}$K). Generally, $Y_{e, \rm eq}$ is smaller for higher density as a result of greater electron degeneracy (which suppresses positron formation and thus favors electron captures).  However, the dependencies differ depending on whether one is in alpha-poor ($\mathcal{G} \gg 1$) or alpha-rich ($\mathcal{G} \ll 1$) regime.  

We generate tables of $Y_{e, \rm eq}(T, \rho)$, $\mu_{\rm eq}(T, \rho)$ and hence $P(T, \rho), U(T, \rho), \Lambda(T, \rho)$ across a grid of temperature and density, which are then inverted to obtain all the needed thermodynamic quantities as functions of density and pressure. This is implemented through the general equation of state module within \texttt{Athena++} \citep{Coleman2020}. 

Our calculations neglect the additional cooling that arises from the dissociation of $\alpha$ particles \citep{Piro&Pfahl07}. We discuss this assumption in Sec.~\ref{sec:alphacool}, and describe those disks for which alpha cooling can compete with, or even dominate, neutrino cooling.  
The temperatures achieved in the marginally Toomre-stable disks of interest are typically only slightly larger than those at which Silicon burning is triggered in the cores of massive stars near the Chandrasekhar mass (e.g., \citealt{Woosley+02}).  Though also neglected in our simulations, thermonuclear burning (e.g., $^{4}$He$+^{16}$O $\rightarrow ^{20}$Ne$+\gamma$), can also occur in the outer regions of collapsar disks, depending on the composition of the infalling star (e.g., \citealt{Zenati+20}) and the effects of this potential energy source on the fragmentation process should also be considered in future work.

\subsection{Initial Conditions}

The disk is initialized to be horizontally uniform, while its vertical structure follows an equilibrium configuration described in \citet{chen2023}.  By specifying an initial midplane density $\rho_0$ and local orbital frequency $\Omega$, we obtain the midplane ``GI factor'',
\begin{equation}
    Q = \dfrac{\Omega^2}{2\pi G \rho_0},
    \label{eq:Q}
\end{equation}
which serves as a rough proxy for Toomre $Q_T$.  One can also define $Q_T$ for a 3D (rather than razor-thin) disk by defining a vertically-averaged sound speed $c_s^2 := P/\rho $ according to:
\begin{equation}
    \left[c_s^2\right]_\rho \equiv \frac{\int_z c_s^2 \rho d z}{\int_z \rho d z},
\end{equation}
such that
\begin{equation}
    Q_T \equiv \frac{\left[c_s^2\right]_\rho^{1 / 2} \Omega}{\pi G \Sigma}.
\end{equation}
When mapping our local simulations to their radial position $r$ around the BH of mass $M_{\bullet}$, we approximate the disk rotation as being Keplerian, for which $\Omega = (GM_{\bullet}/r^{3})^{1/2}$.

For any self-similar, barotropic vertical profile, $Q_T$ becomes a function of $Q$ only (for details, see \citealt{chen2023}). The disk's vertical density profile $\rho(z)$ is set by hydrostatic equilibrium,
\begin{equation}
\frac{1}{\rho} \frac{d P}{d z}=-\Omega^2 z-4 \pi G \int_0^z \rho\left(z^{\prime}\right) d z^{\prime}.
\label{eqn:hydrostatic}
\end{equation}
We follow \citet{chen2023} and take $P \propto \rho^{\gamma_0}$ for $\gamma_0 = 4/3$, so chosen that radiation pressure support does not diverge unphysically at high $z$. 
Because the disk's structure will eventually either settle into a quasi-stationary profile (for long cooling times), or undergo runaway fragmentation and collapse (short cooling time), our results are not sensitive to the value of $\gamma_0$. 

The midplane density $\rho_0 = \rho(z= 0)$ in the initial state follows from the assumed values of $Q_T = 1.0$ (equivalently, $Q = 0.76$) and $\Omega$. The only remaining physical scale to be determined is the midplane temperature $T_0$, or equivalently pressure $P_0(T_0, \rho_0)$, which, once chosen, 
sets the entire vertical profile.  The scale height  $H = \sqrt{P_0/\rho_0} / \Omega$  determines the characteristic surface density unit $\Sigma  = 2 \rho_0 H =  2 \sqrt{P_0 \rho_0} / \Omega$ which differs by only an order unity factor ($\sim 0.8$) from the exact integration of $2 \int_z \rho d z$. 

We initialize a decaying turbulence field using the turbulence module of \texttt{Athena++}, which distributes kinetic energy across wavenumber 1 to 16 with a spectral slope that we choose to be -2, similar to \citet{chen2023}. Once a long-term steady state is achieved, the outcome is not expected to depend on the details of the initial turbulence field.  We have confirmed this convergence by testing representative fragmentating simulations for different turbulence assumptions.

\subsection{Boundaries and Resolution}
The default box size for our simulations is $(6H)^3$, 
resolved with a grid of $256^3$. 
For both hydrodynamic variables and gravity we apply the standard shearing-periodic boundary condition in $x$ and periodic boundary condition in $y$. 
We implement an open outflow boundary condition in $z$ by setting the density and pressure in the boundary cells to the same values as the last active cells. We do the same for velocity if directed outwards and set it to 0 if otherwise.  The Poisson solver for gravity applies a vacuum boundary conditions along $z$. The other conditions we apply are similar to \citet{chen2023} but without the added complications from radiative transfer. We implement floors on $P$ and $\rho$ such that they do not fall below $10^{-3}$ of their initial midplane values anywhere on the grid.  Our floor values are sufficiently small that in most of simulations we were able to prevent significant mass loss from outflows at the vertical boundaries and achieve mass conservation.  Table \ref{tab:para} summarizes our suite of simulations, which span a range of $\rho_0$ and $\Omega$ to cover marginally-stable $Q_T \sim 1$ disks with different properties relevant to collapsar disks.

\subsection{Identifying Bound Objects}
\label{sec:diagnostics}

For disks that undergo gravitational collapse, 
we halt the simulation once clump cores become dense enough such that the central gravitational timestep drops sharply, and the Riemann solver fails, which occurs typically after of 1-2 initial cooling timescales. 
At this end state, 
we identify gravitationally bound objects using the method of \citet{chen2023}, 
building on the approach of \citet{Mao2020}.
For each local minimum in the gravitational potential field $\Phi$, we first identify the largest closed contour 
that contains no other local minima $\Phi = \Phi_{\rm max}$. 
Within this region of interest, we narrow the region down to cells where total kinetic and thermal energy is smaller than total gravitational potential with respect to $\Phi_{\rm max}$ to generate genuinely bound structures. Within the bound structures (``clumps''), we calculate $M_{\rm clump}$ and $Y_{e, \rm clump}$ 
as the total mass and 
average electron fraction 
within the clump to determine whether each clump is likely to continue to collapse to NS by comparing $M_{\rm clump}$ to $M_{\rm Ch}(Y_{\rm e,clump})$ (Eq.~\eqref{eq:MCh}). The clump masses are typically of order the Jeans mass (an estimate of which is given below in Eq.~\eqref{eq:MJ}).

\section{Results}
\label{sec:results}

We begin in Sec.~\ref{sec:general_remarks} with an overview of our simulation results, before describing separately those models that achieve a quasi-steady turbulence accretion state (Sec.~\ref{sec:steady}) versus those that fragment and undergo gravitational collapse (Sec.~\ref{sec:fragment}).

\begin{table*}
\centering
  \begin{tabular}{c|ccccccccc}
     \hline\hline
     Model & $\Omega$ & $\rho_0$  & $T_0$[K]  &  $\Sigma$  & $M_{\rm d}(<r)^{(a)}$ & $\tau_{\rm cool}$ & H & Outcome & Figure \\
     \hline
Unit &      [rad s$^{-1}$] & [g/cm$^3$] & [K] & [g/cm$^2$] & [$M_{\odot}$] & - & [km] & - & - \\
     \hline
        \texttt{O16T1e10} & 16.0 & $8.0 \times 10^8$ &  $1.0\times 10^{10}$& $6.19\times 10^{16}$ & 1.92 & 3.08 & 413 & Fragment & \ref{fig:MchvsMclump_all} \\
        \texttt{O16T1.5e10} & ... & ... & $1.5\times 10^{10}$&  $1.14\times 10^{17}$ & 2.40 &  3.97& 707 & Fragment &    \ref{fig:frag_vertsnap},  \ref{fig:frag_midsnap}, \ref{fig:MchvsMclump},  \ref{fig:Xalphahighsigma}\\
        \texttt{O12T1e10} & 12.0 & $4.5\times 10^8$ & $1.0\times 10^{10}$&  $5.12\times 10^{16}$ &1.59 & 4.43 &568 & Fragment &  \ref{fig:MchvsMclump_all} \\
        \texttt{O12T1.5e10} & ... & ... & $1.5\times 10^{10}$ &$9.41 \times 10^{16}$ & 2.92 & 3.51 & 1044  & Fragment &   \ref{fig:MchvsMclump_all}\\
        \texttt{O8T1e10} & 8.0 & $2.0\times 10^8$ & $1.0\times 10^{10}$  & $3.85 \times 10^{16}$ &2.07 & 7.20 & 963 & Fragment &    \ref{fig:alpha_frag_midsnap},  \ref{fig:Xalphahighsigma},  \ref{fig:MchvsMclump_alpha} \\
        \texttt{O8T1.5e10} & -- & -- & $1.5\times 10^{10}$&  $8.65\times 10^{16}$ & 4.06 & 3.33 & 1897 & Fragment &\ref{fig:MchvsMclump_all}\\
        \texttt{O8T5e9} & -- & -- & $5\times 10^{9}$&  $2.21\times 10^{16}$ &1.27 & 23.50 & 599 & Quasi-steady & \ref{fig:turbulent_vertsnap}, \ref{fig:turbulent_midsnap}\\
        \hline
        \hline
        \end{tabular}
   \caption{Summary of our simulation suite of marginally gravitationally stable disks. Most models assume $Q_{\rm T} = 1$ in the initial state.  ... means ``as above.'' $^{(a)}$ Local disk mass $M_{\rm d}(<r) \equiv \pi \Sigma r^{2}$, where the disk radius is calculated as $r = (GM_{\bullet}/\Omega^{2})^{1/3}$ for $M_{\bullet} = 3M_{\odot}$.}
   \label{tab:para}
  \end{table*}

\begin{figure*}[htbp]
\centering
\includegraphics[width=0.8\textwidth,clip=true]{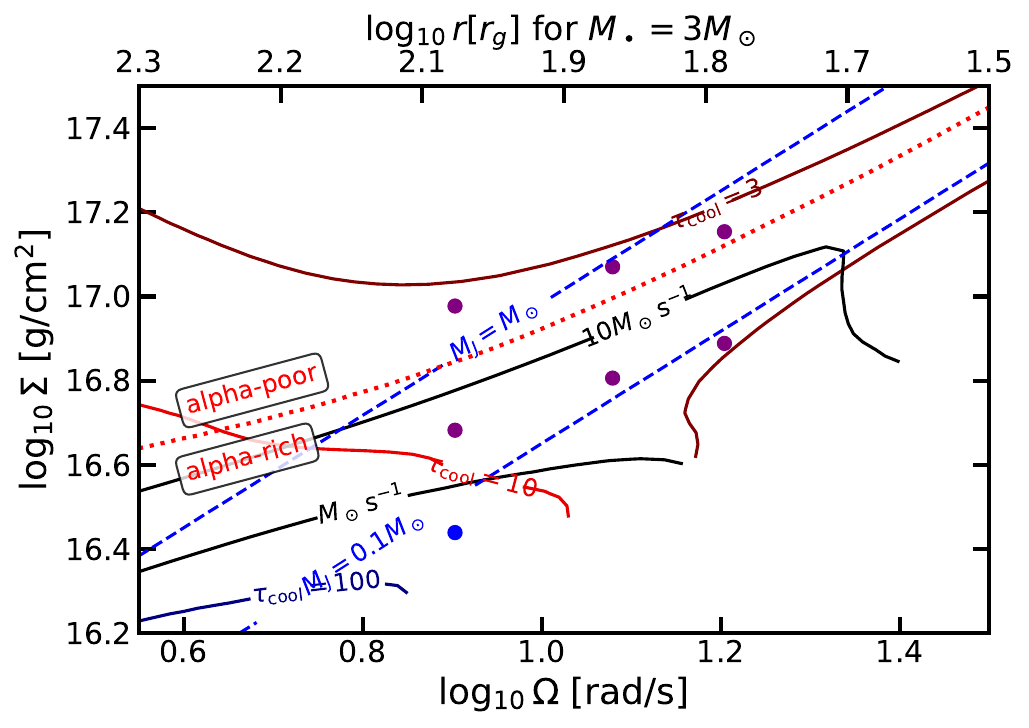}
\caption{Locations of our suite of disk models in the space of surface density $\Sigma$ and disk angular velocity $\Omega$. Purple circles indicate those models that undergo runaway cooling and fragmentation (Sec.~\ref{sec:fragment}), 
while a blue symbol denotes the single simulation that evolved into a quasi-steady state supported by gravito-turbulence (Sec.~\ref{sec:steady}). 
Shown for comparison with solid colored lines are contours of the dimensionless neutrino cooling timescale $\tau_{\rm cool, 0}$. We see that $\tau_{\rm cool, 0} \lesssim 10$ (below the red solid line) delineates our fragmenting models. 
Also shown are contours of the local mass accretion rate $\dot{M}$ (Eq.~\eqref{eq:Mdot}; black solid lines) and local clump mass unit $\sim $ Jeans mass in solar masses (Eq.~\eqref{eq:MJ}; blue dashed lines). Along the top horizontal axis we show the disk radius in gravitational radii $r_g \equiv GM_{\bullet}/c^{2}$ for a central BH of mass $M_{\bullet} = 3M_{\odot}$.
}
\label{fig:contours_tcool}
\end{figure*}

\subsection{General Remarks}
\label{sec:general_remarks}

Drawing on insights from simulations of gravitational instability in other contexts \citep{Johnson2003, Jiang2011, chen2023},
we parameterize our models according to their initial cooling time $t_{\rm cool,0} = U(\rho_0, P_0)\Omega/\Lambda(\rho_0, P_0)$ (Fig.~\ref{fig:tcool}) normalized to the orbital frequency according to $\tau_{\rm cool, 0} \equiv t_{\rm cool,0}\Omega$. Assuming $Q_T = 1$, $\tau_{\rm cool, 0}$ is determined by $\Sigma$ and $\Omega$, whose relationships to $\rho_0$ and $P_0$ were described in Sec.~\ref{sec:setup}. Runaway cooling to form gravitationally bound clumps is expected when $\tau_{\rm cool} \lesssim 5$–10 \citep{Gammie2001,Lodato2004,Rice+05}, a threshold that becomes higher when the disk is radiation dominated \citep{chen2023}.

Figure \ref{fig:contours_tcool} shows the locations of our simulation suite (Table \ref{tab:para}) in the $(\Sigma, \Omega)$ parameter space.  
For a given BH mass, $\Omega = (GM_{\bullet}/r^{3})^{1/2}$ may be taken as a proxy for the local disk radius $r$ from the BH (shown in gravitational radii for $M_{\bullet} = 3M_{\odot}$ along the top axis).  Purple circles denote those models that undergo fragmentation, 
while a blue circle marks the single case that evolves into a state of quasi-steady turbulence as a control simulation. 
Colored contours of the initial cooling timescale, $\tau_{\rm cool, 0} = 3, 10, 100$, which illustrate that the models with $\tau_{\rm cool, 0} \lesssim 10$ (large $\Omega$) fragment, while our quasi-steady model has larger initial $\tau_{\rm cool,0}\sim 20$.

A red dashed line shows the $\mathcal{G}(\rho_0, T_0) =1$ contour (Eq.~\eqref{eqn:Gmaintext}, Fig.~\ref{fig:Yeeq-rho-T}), separating disks with large $\alpha$ particle mass fractions from those in which free nucleons dominate (Appendix~\ref{app:EOS}).  
For models with $\Omega < 160$ rad s$^{-1}$ below the red dashed line, the disk composition is dominated by $\alpha$ particles in the initial state (i.e, $X_{\alpha} \simeq 1$).  As we shall discuss, several of these disks also fragment, and the $\alpha$ particles are dissociated within the collapsing clumps.

Our local simulations can be approximately mapped into the global disk properties by calculating the characteristic ``accretion rate'' for each model, 
\begin{equation}
    \dot{M} \approx 3\pi \alpha \dfrac{(\pi G)^2 \Sigma^3}{\Omega^3} \sim 3\pi  \dfrac{(\pi G)^2 \Sigma^3}{\Omega^3 \tau_{\rm cool}},
    \label{eq:Mdot}
\end{equation}
which, for simplicity, is defined exclusively based on the local properties in the $Q_T \approx 1 $ initial state. 
In the second line we use $\alpha \sim \tau_{\rm cool}^{-1}$ \citep{Gammie2001} as the maximum effective viscosity that can be supplied by gravitoturbulence (e.g., \citealt{Gammie2001}).

Roughly speaking, $\dot{M}$ is the rate that gas must flow into the local radial annulus, via direct infall from the collapsing star or through disk accretion from larger radii, to maintain the marginally-stable $Q_T \approx 1$ initial state.  In this sense $\dot{M}$ only represents the true radial accretion rate through the disk \textit{if} it is supported self-consistently via gravito-turbulent heating, which our simulations indicate are unlikely across much of the relevant parameter space because the disk fragments.\footnote{Or unless additional heating mechanisms operate apart from disk turbulence, such as feedback from accretion onto the formed compact objects (e.g., \citealt{Lerner2025}).} We have confirmed that our non-fragmenting simulations yield a quasi-steady accretion rate to within a factor of order unity of this estimate. 

Contours of $\dot{M}$ are shown with black lines in Fig.~\ref{fig:contours_tcool}.
For $\Omega < 20 $ rad s$^{-1}$ the $\dot{M} = 10M_\odot$ s$^{-1}$ contour roughly traces the $\mathcal{G}\approx 1$ contour, before the two contours deviate at larger $\Omega$. 
Thus, for disks with an initial composition dominated by free nucleons ($X_\alpha \ll 1$; $\mathcal{G} \gg 1$) to fragment requires very high accretion rates $\dot{M} \gtrsim 10 M_\odot \mathrm{s}^{-1}$, which to be achieved may require the collapse of very massive stars $\gtrsim 100M_{\odot}$ (e.g., \citealt{Siegel+22}). 
However, as already mentioned and described in detail below, alpha-rich disks ($\mathcal{G} \ll 1$) can also cool efficiently and fragment ($\tau_{\rm cool,0} < 3-10$) even for somewhat lower accretion rates $\dot{M}\sim M_{\odot}$ s$^{-1}$, closer to those achieved by the collapse of more typical progenitor stars (e.g., \citealt{MacFadyen&Woosley99}).

The marginally stable $Q_T = 1$ initial disks modeled here do not necessarily apply to arbitrarily small radii (high $\Omega$) close to the BH. Height-integrated one-dimensional disk models that assume a radially-constant accretion rate $\dot{M} \gtrsim M_{\odot}$ s$^{-1}$ (e.g., \citealt{ChenBel2007,Piro&Pfahl07}; \citetalias{metzger2024}), predict that only the outer regions of the disk $\gtrsim 100 R_{\rm g}$ become gravitationally unstable.  At smaller radii, other sources of turbulence, such as that driven by the magneto-rotational instability (MRI; \citealt{Balbus&Hawley98}), provide sufficient viscosity ($\alpha_{\rm NDAF} \sim 0.1$) to maintain a neutrino-cooled disk with $Q_T > 1$. 

A notable feature of self-gravitating neutrino-cooled disks is that the dimensionless cooling timescale generally {\it increases} with $\Omega$.  
This implies that the innermost radius where self-gravity first becomes important ($Q_T \sim 1$) are the most unstable to collapse, with $\tau_{\rm cool}\sim 1/\alpha_{\rm NDAF}$.\footnote{This behavior contrasts with self-gravitating AGN disks, for which the cooling timescale is typically smaller at large radii (small $\Omega$), rendering their outer regions more susceptible to fragmentation (e.g., \citealt{chen2023}).}

For disks that fragment, the resulting clumps possess a characteristic mass scale set by the Jeans mass (also the unit mass in our code),
\begin{equation}
    M_{\rm J} \equiv \rho H^{3} \simeq \frac{M_{\bullet}}{2\pi} \left(\frac{H}{r}\right)^{3} \simeq 0.02M_{\odot}\left(\frac{M_{\bullet}}{3M_{\odot}}\right)\left(\frac{3H}{r}\right)^{3},
    \label{eq:MJ}
\end{equation}
where we have used Eq.~\eqref{eq:Q} for $Q = 1$. 
Given the expected vertical aspect ratio of collapsar disks, $H/r \sim 0.2-1$ \citep{ChenBel2007}, we find values $M_{\rm J} \sim 10^{-3}-1M_{\odot}$ in the sub-solar mass regime when mapped to a global parameter space (e.g., \citetalias{metzger2024}; \citealt{Lerner2025}). 
Blue dashed lines in Fig.~\ref{fig:contours_tcool} show contours of $M_{\rm J}$ across the model parameter space for $M_{\bullet} = 3M_{\odot}$,  revealing $M_{\rm J} \sim 0.1-1M_{\odot}$ for our fragmenting disk models.

\subsection{Quasi-Steady Turbulence}
\label{sec:steady}

\begin{figure*}[htbp]
\centering
\includegraphics[width=1.02\textwidth,clip=true]{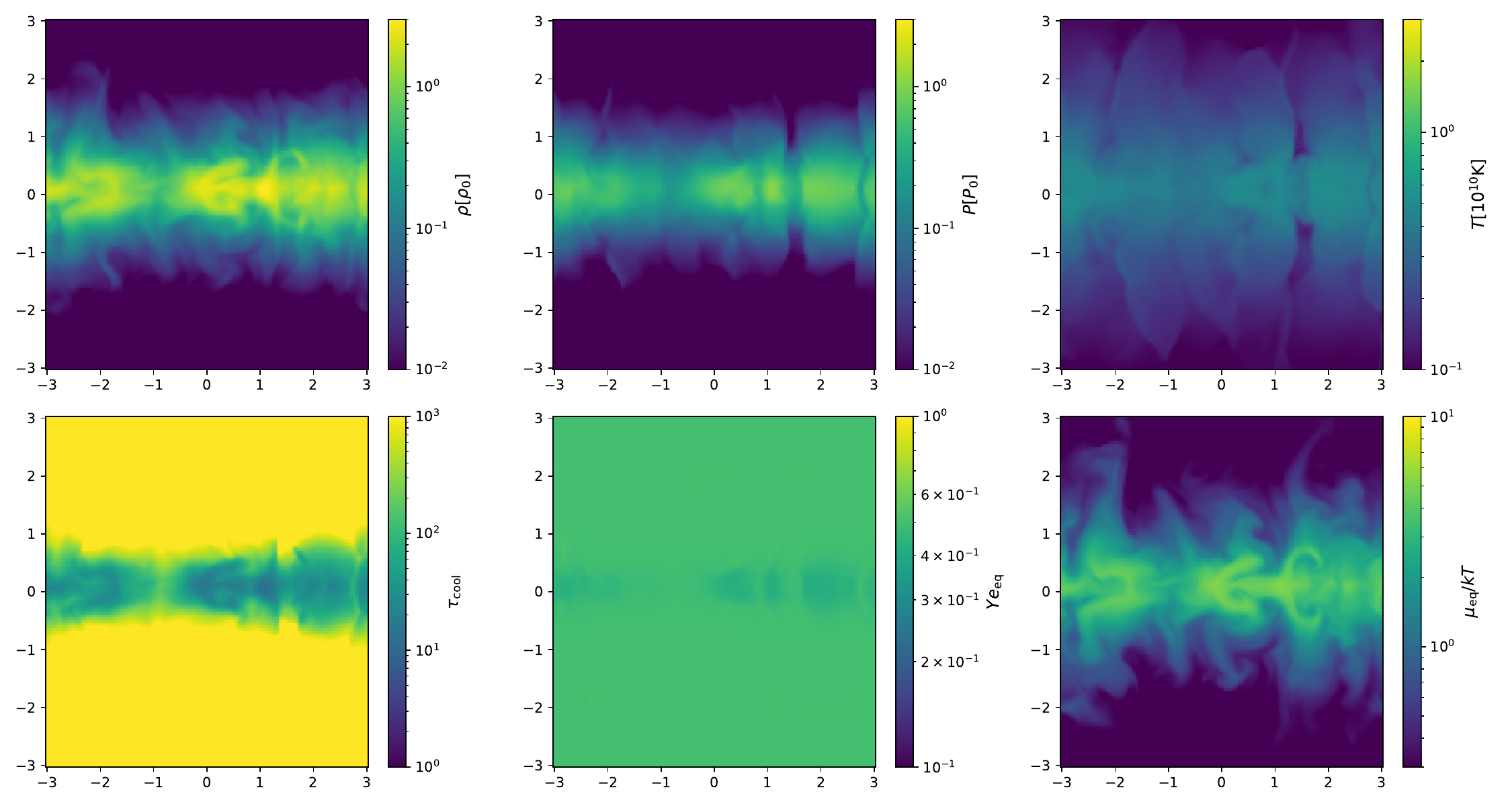}
\caption{For the fiducial slow cooling (non-fragmenting) simulation, a typical snapshot showing vertical $(x, z)$ profiles through the $y = 0$ plane of density, temperature, local cooling time, equilibrium electron fraction and chemical potential. Lengths are expressed in units of the vertical scale-height $H$.}
\label{fig:turbulent_vertsnap}
\end{figure*}

\begin{figure*}[htbp]
\centering
\includegraphics[width=1.02\textwidth,clip=true]{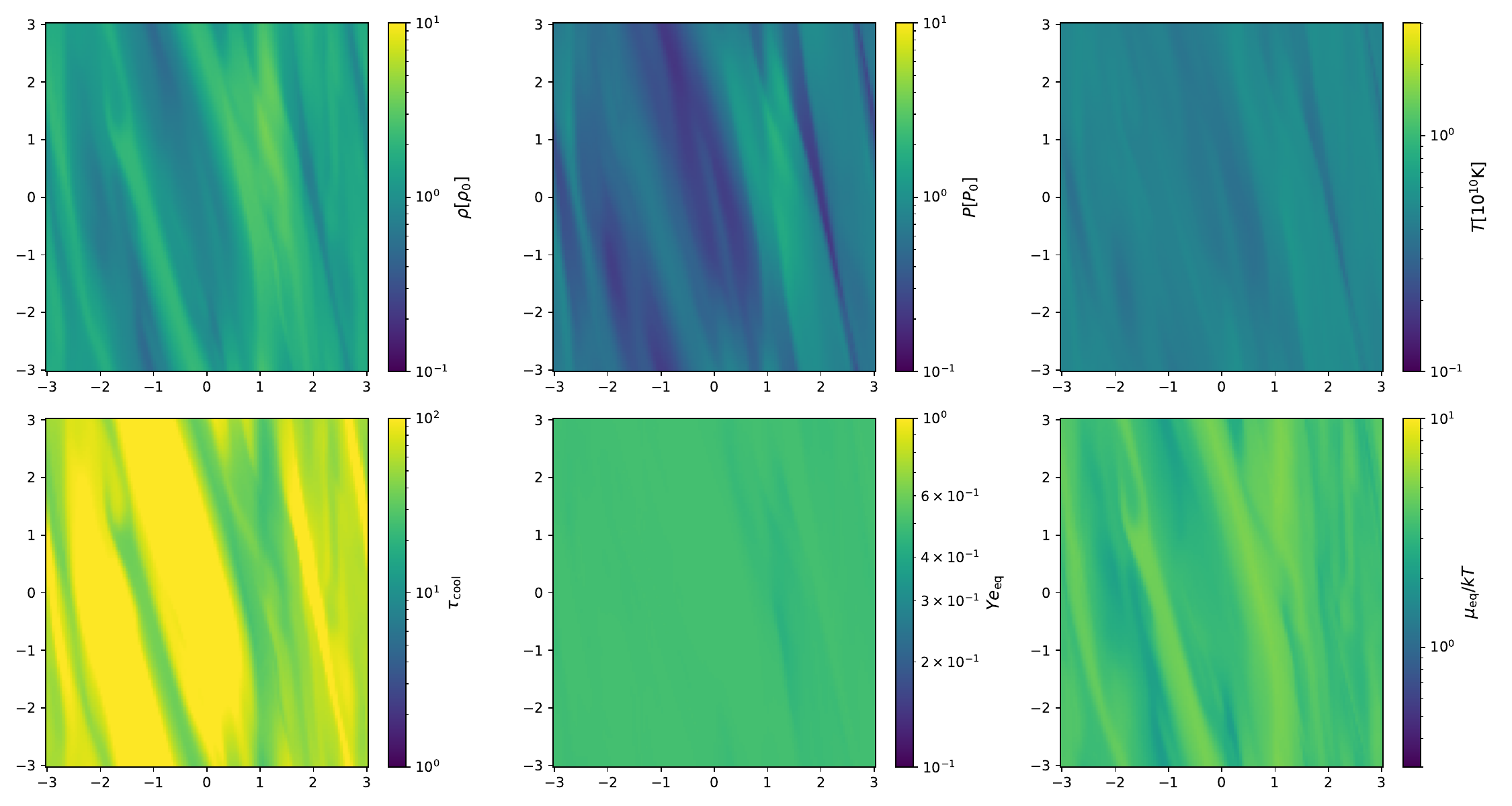}
\caption{For the fiducial slow cooling (non-fragmenting) simulation, a typical snapshot showing the $(x, y)$ profiles through the midplane ($z = 0)$ of density, temperature, local cooling time, equilibrium electron fraction and chemical potential. Lengths are expressed in units of the vertical scale-height $H$.}
\label{fig:turbulent_midsnap}
\end{figure*}

To explore the limits of the parameter space, and as a check on our simulations, we first confirm that for sufficiently long cooling timescales (large $\tau_{\rm cool,0}$), the disk finds a quasi-steady gravito-turbulent state. This turbulence generates enough heating to maintain the disk in a marginally-Toomre unstable state $Q \sim 1$ with time-averaged properties similar to the assumed initial state.  For our stable simulation \texttt{O8T5e9}, Figs.~\ref{fig:turbulent_vertsnap}, \ref{fig:turbulent_midsnap} show snapshot slices through the vertical ($y = 0$) and midplane ($z = 0$), respectively.  Although the midplane is mildly degenerate $\mu/kT \sim \mathcal{O}(1)$, $Y_e$ remains close to a half because the density and temperature are sufficiently low that $\alpha$ particles dominate the disk composition.  The cooling is strongest in compact density wave structures close to the disk midplane, where $\tau_{\rm cool}\sim 40$ locally. Although these dense structures dominate the overall cooling rate of the disk, cooling is negligible throughout the bulk of its volume ($\tau_{\rm cool} > 10^3$). Mild turbulence generates enough heating to balance the radiative cooling, leading to an effective local viscous heating rate $\alpha_{\rm GI} \sim \tau_{\rm cool}^{-1}$ \citep{Gammie2001}.

\subsection{Fragmenting Disks}
\label{sec:fragment}

Our models with short cooling times undergo fragmentation.  
In these cases, over a few cooling timescales, 
the peak density and pressure in the midplane rapidly increase and $Q_T$ drops as turbulent heating fails to prevent the disk from collapsing.

\subsubsection{High density, initially alpha-poor disks ($X_\alpha \ll 1, Y_{e} <  0.5)$}
\label{sec:highdensity}

\begin{figure*}[htbp]
\centering
\includegraphics[width=1.02\textwidth,clip=true]{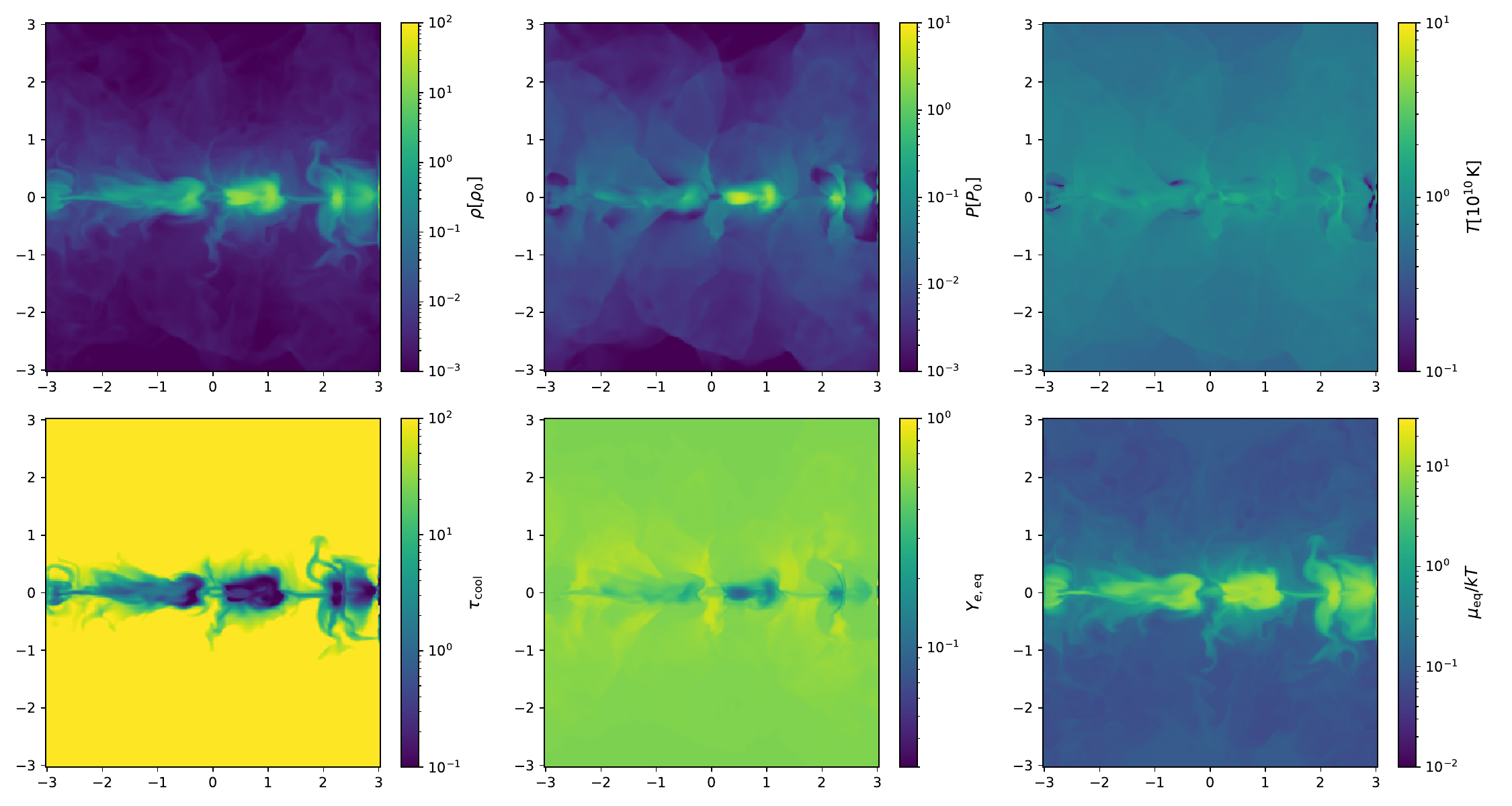}
\caption{Final snapshot of the vertical $(x,z)$ structure of density, 
temperature, dimensionless cooling time, electron fraction $Y_{e}$ and electron degeneracy parameter $\mu/kT$ taken along a constant $y = -3 $ slice from the fragmenting simulation \texttt{O16T1.5e10}. Lengths are expressed in units of the vertical scale-height $H$.}
\label{fig:frag_vertsnap}
\end{figure*}

\begin{figure*}[htbp]
\centering
\includegraphics[width=1.02\textwidth,clip=true]{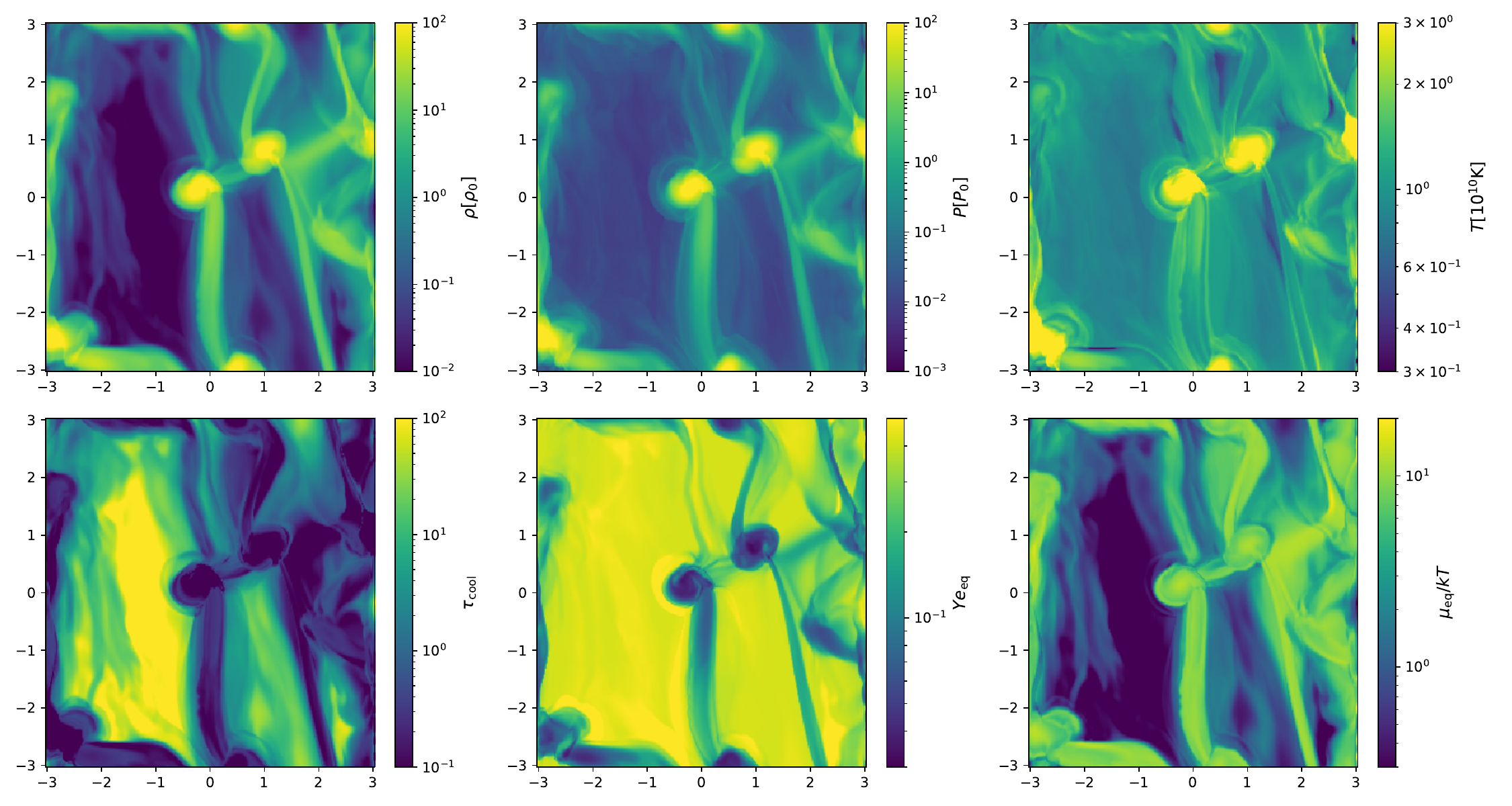}
\caption{Similar to Fig.~\ref{fig:frag_vertsnap} but showing a slice through the midplane from the end of the fragmenting simulation \texttt{O16T1.5e10}. Lengths are expressed in units of the vertical scale-height $H$. }
\label{fig:frag_midsnap}
\end{figure*}

When $\Sigma$ (or, equivalently, $\dot{M}$) is high, the fragmenting disk is dominated by free nucleons in its initial state ($\mathcal{G} \gg 1$). One example is model \texttt{O16T1.5e10} located in the upper right hand corner of Fig.~\ref{fig:contours_tcool}. Snapshots taken at the end of this simulation of the vertical (Figure \ref{fig:frag_vertsnap}) and midplane (Figure \ref{fig:frag_midsnap}) structure reveal the formation of dense bound clumps with 
$Y_{e} \lesssim 0.1$, corresponding to a low effective Chandrasekar mass in these regions $M_{\rm Ch} \lesssim 0.05M_{\odot}$.  These clumps also correspond to local maxima in the pressure, which is dominated by the contribution from non-relativistic neutrons (the small value of $Y_{e}$ limits the pressure contribution from electrons and the high degeneracy limits the positron density; see Fig.~\ref{fig:tcool}, left panel). 

\begin{figure}[htbp]
\centering
\includegraphics[width=0.49\textwidth,clip=true]{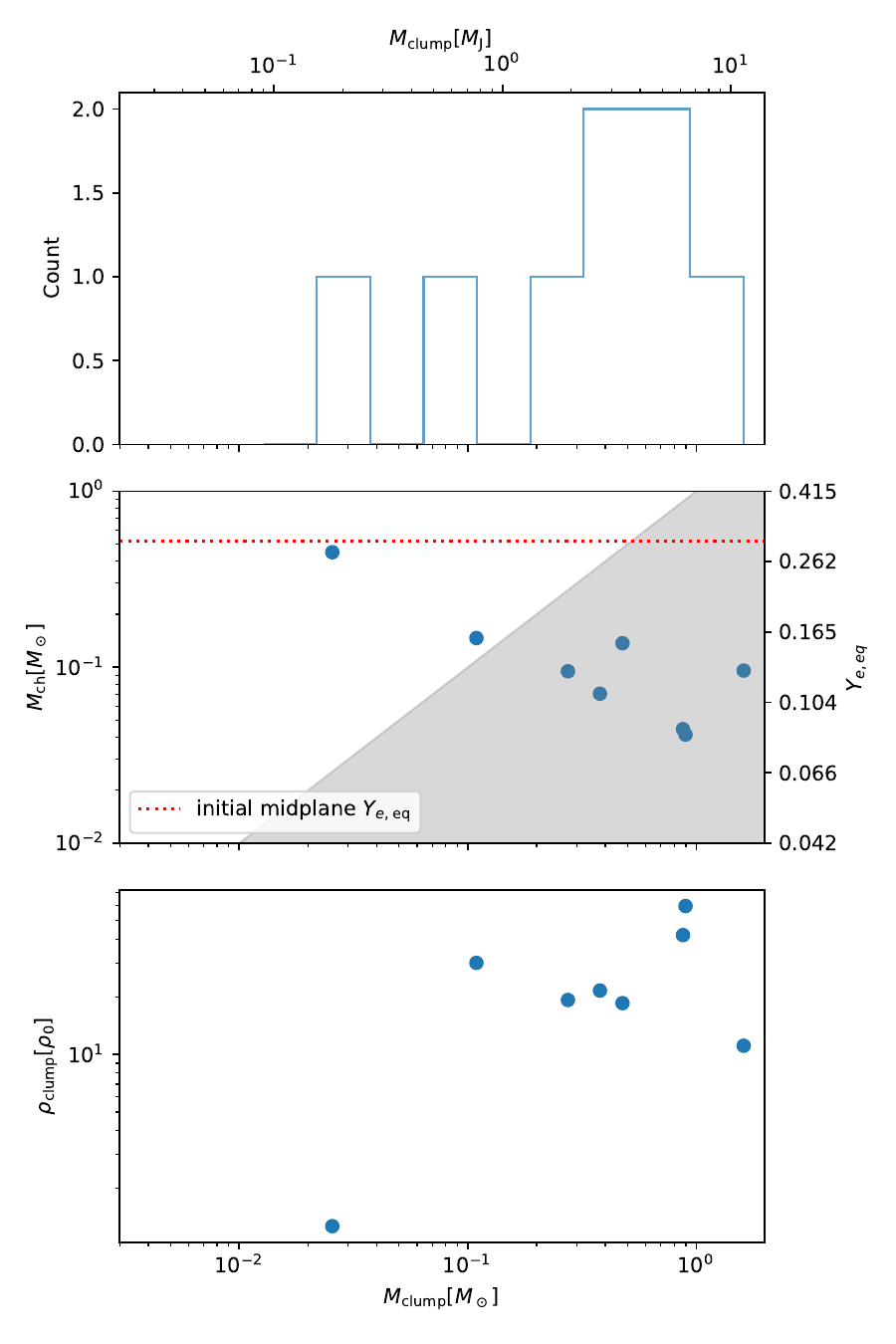}
\caption{Properties of the self-bound clumps from the fragmenting simulation \texttt{O16T1.5e10}. Upper panel: Mass histogram; Middle panel: Average $Y_e$ (left axis) and associated Chandrasekhar mass $M_{\rm Ch}(Y_{e})$ (right axis; Eq.~\eqref{eq:MCh}). Clumps in the gray shaded region have masses $\gtrsim M_{\rm Ch}$ and hence are expected to continue to collapse to form NSs.  Lower panel: average density in units of the initial midplane density. }
\label{fig:MchvsMclump}
\end{figure}

At the final snapshots shown above, we measure several properties of the bound clumps. 
These are shown in Figure \ref{fig:MchvsMclump} as a function of the clump mass expressed both in terms of the physical mass (bottom axis) and the Jeans mass $M_{\rm J} = \rho_0 H^3$ (top axis; Eq.~\eqref{eq:MJ}).  
The histogram of the clump masses shown in top panel of Figure \ref{fig:MchvsMclump} indicates that most are within a factor of a few of the local Jeans mass.  From the average electron fraction (right axis) and associated Chandrasekhar mass (left axis) of the clumps shown in the middle panel, we see that typically $M_{\rm clump} > M_{\rm Ch}$.  Most of the clumps are thus likely to continue to collapse to higher densities, eventually forming low-mass NSs within seconds or less. However, following this process explicitly would require far greater dynamic range than captured by our simulations (the mean density of the final NS exceeds that of the clumps reached in our simulations by 3-4 orders of magnitude), as well as the inclusion of neutrino transport effects. As the clumps contract, their electron fractions drop by a significant factor $\sim 3$ due to electron captures; the latter dominate positron captures due to the higher densities (higher electron degeneracy) of the clumps compared to the original disk midplane density $\rho_0$ (bottom panel of Fig.~\ref{fig:MchvsMclump}). 

We terminate our simulations once the timestep becomes very short as the clumps undergo runaway collapse (both the dynamical and cooling timescales become short at higher density).  By this point, we see no evidence that the clumps have merged with each other.  Nevertheless, such mergers may take place over longer timescales, 
and it is of interest to know whether this will occur frequently before the clumps have reached their final state as NSs. 
To this end, we compare the free-fall time of the clumps, $t_{\rm ff} = 1/\sqrt{G \rho_{\rm clump}}$, to a rough estimate of their dynamical merger timescale,

\begin{equation}
    t_{\rm merge} \approx (\sigma R_{\rm clump} v_{\rm merge})^{-1},
    \label{eqn:merge}
\end{equation}
where $\sigma$ is the surface number density of clumps near the midplane and $R_{\rm clump} \sim (M_{\rm clump}/\rho_{\rm clump})^{1/3}$ is an estimate of the average clump size. Taking a typical relative velocity between clumps to be set by the disk's shear $v_{\rm merge} \sim \Omega R_{\rm clump}$, we find that

\begin{equation}
\dfrac{t_{\rm ff}}{t_{\rm merge}}  =  \sigma H^2  \dfrac{(M_{\rm clump}/M_{\rm J})^{2/3}}{(\rho_{\rm clump}/\rho_0)^{7/6}}
\label{eq:tff2tmerge}
\end{equation}

The distributions of $M_{\rm clump}$ and $\rho_{\rm clump}$ shown in Figure \ref{fig:MchvsMclump} reveal this ratio is typically $t_{\rm ff}/t_{\rm merge} < 1$.  Insofar that $M_{\rm clump} \gtrsim M_{\rm J}$, and the clumps are area-filling $\sigma H^2 \sim 1$, this conclusion mainly follows from the high density contrast of the clumps, $\rho_{\rm clump}/\rho_0 \gtrsim 10$.  Nevertheless, the fact that $t_{\rm ff}$ and $t_{\rm merge}$ are comparable to order of magnitude suggests that mergers, or the formation of bound binaries from the collapsed NS products, may be expected to occur relatively shortly after NS formation (Appendix ~\ref{sec:GWmerger}).

\subsubsection{Low density, initially alpha-rich disks ($X_\alpha\sim 1, Y_{e}\sim 0.5$)}

\label{sec:lowdensity}

Lower density disks, corresponding to models located below the $\mathcal{G}(\rho, T) = 1$ line in Fig.~\ref{fig:contours_tcool}, typically start their evolution with higher initial $Y_{e} \approx 0.5 $ and $X_{\alpha}\approx 1$ in the midplane.  These solutions correspond to somewhat lower accretion/infall rates $\dot{M}\sim M_\odot$ s$^{-1}$ than the model presented in the previous section (with $\mathcal{G}(\rho, T) > 1$ and lower initial $Y_{e}$) and hence are more readily achieved in collapsars.

\begin{figure*}[htbp]
\centering
\includegraphics[width=1.02\textwidth,clip=true]{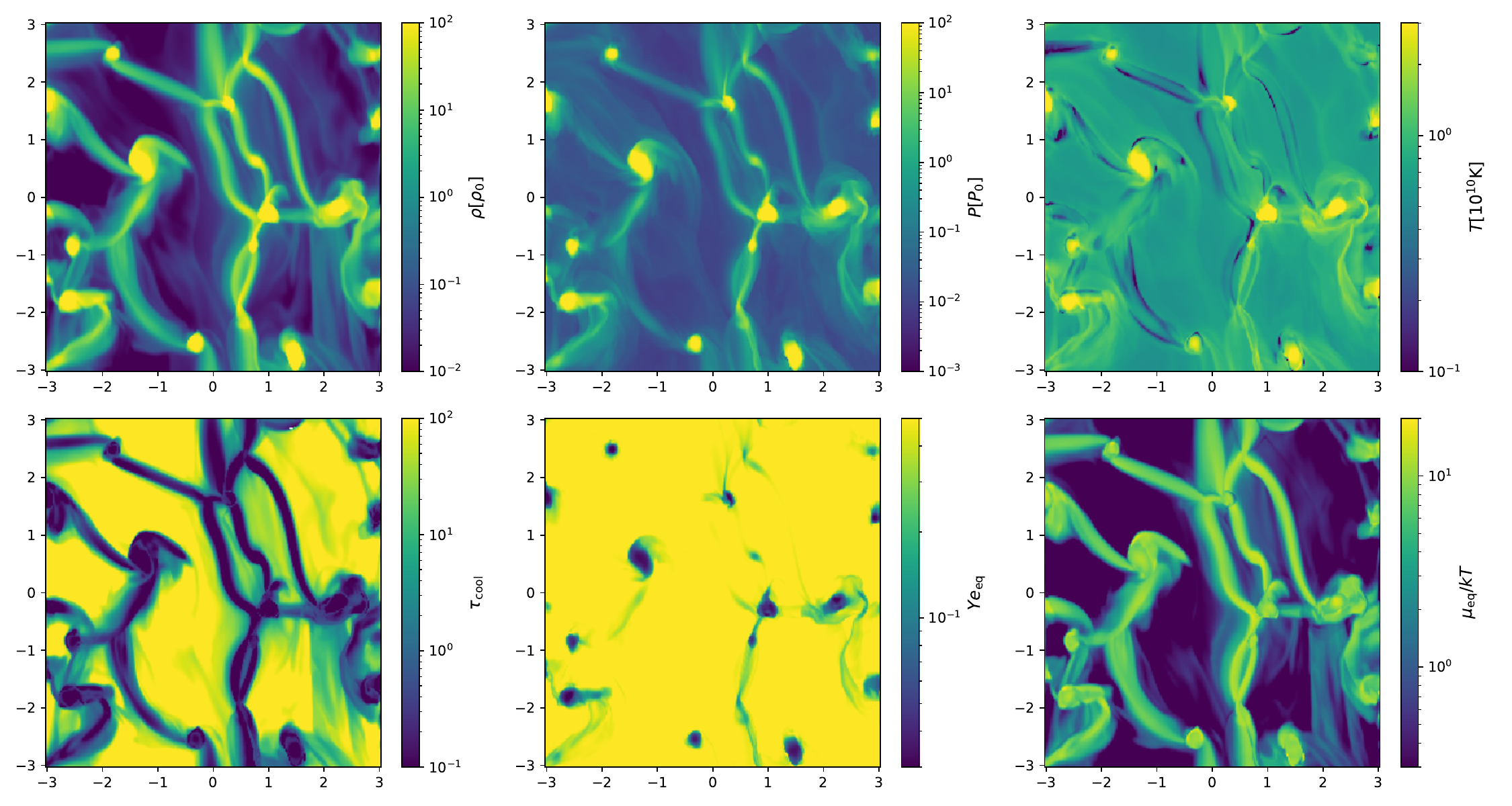}
\caption{Snapshot of the $(x,y)$ profiles through the midplane ($z = 0$) of density, temperature, dimensionless cooling time, electron fraction $Y_{e}$ and $\mu/kT$ for the fragmenting simulation \texttt{O8T1e10}. Lengths are expressed in units of the vertical scale-height $H$.}
\label{fig:alpha_frag_midsnap}
\end{figure*}

Although $Y_{\rm e}$ starts close to 0.5 for these disks, this does not mean they can only fragment to massive NS $\gtrsim M_{\rm Ch}(Y_{e} = 0.5) \sim 1.4M_{\odot}.$  As clumps undergo runaway cooling and contract to high densities (e.g., Fig.~\ref{fig:MchvsMclump}) both $Y_{e}$ and $X_{\alpha} \simeq 2Y_{e}$ decrease.  Figure \ref{fig:alpha_frag_midsnap} shows a snapshot of a slice through the midplane for model \texttt{O8T1e10}. Although the bulk of the midplane is seen to retain $Y_e \approx 0.5$, the collapsing clumps achieve $Y_e \ll 0.5$ and hence can form deeply sub-solar bodies ($M_{\rm Ch} \ll M_{\odot})$ similar to the higher $\dot{M}$ case.

\begin{figure*}[htbp]
\centering
\includegraphics[width=1.1\textwidth,clip=true]{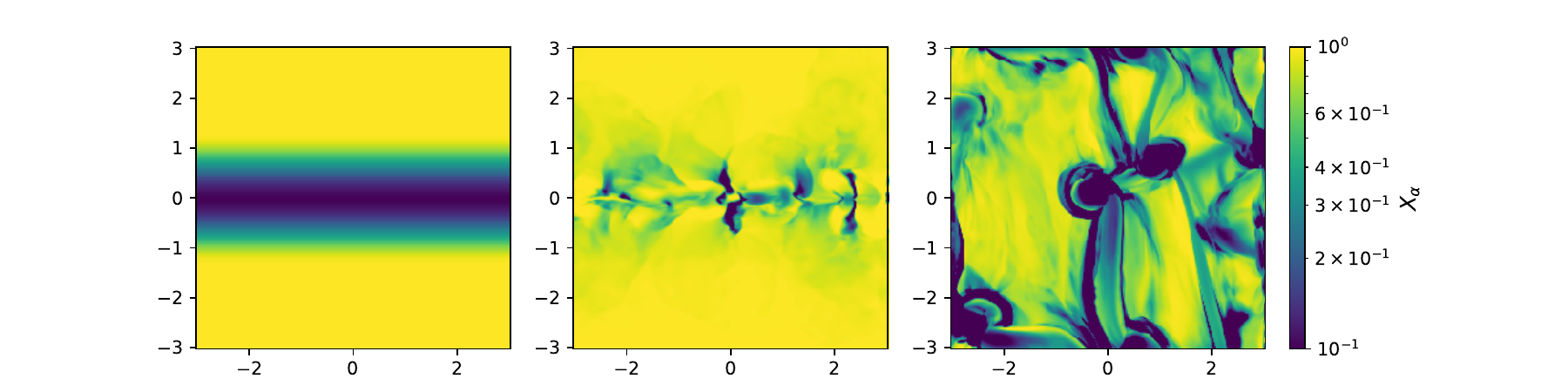}
\includegraphics[width=1.1\textwidth,clip=true]{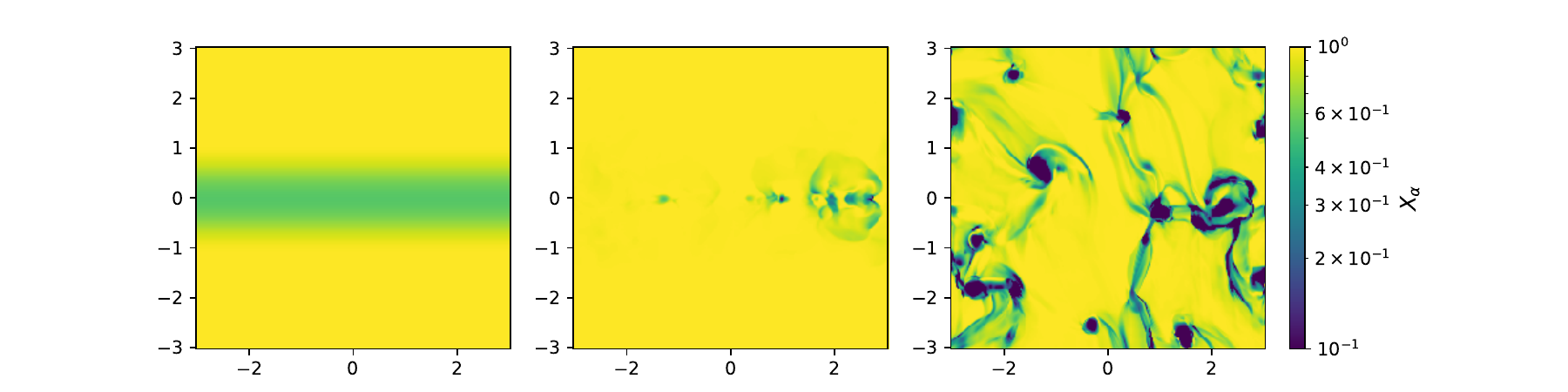}
\caption{Snapshots show the vertical profiles of $X_\alpha$ through a $y = {\rm Const}$ slice comparing the initial state to that after fragmentation, 
as well as the $X_\alpha$ distribution in the midplane in the fragmentation state. 
The upper and lower panels are for case \texttt{O16T1.5e10} and case \texttt{O8T1e10}, respectively. }
\label{fig:Xalphahighsigma}
\end{figure*}

Figure \ref{fig:Xalphahighsigma} compares snapshots of the vertical (initial and after fragmentation) and midplane (after fragmentation) structure of 
$X_\alpha$ of the same model \texttt{O8T1e10} (lower panel) with the higher-$\dot{M}$ model \texttt{O16T1.5e10} presented in Sec.~\ref{sec:highdensity} (upper panel).  This more explicitly shows that \texttt{O8T1e10} exhibits a more alpha-rich $X_\alpha \approx 1$ initial midplane than \texttt{O16T1.5e10}. However, as clumps in the disk reaches densities $\sim 10-100$ times greater than the background their $\alpha$ particles are destroyed (even though $X_\alpha \approx 1$ 
is maintained in the surrounding less dense regions).

\begin{figure}[htbp]
\centering
\includegraphics[width=0.49\textwidth,clip=true]{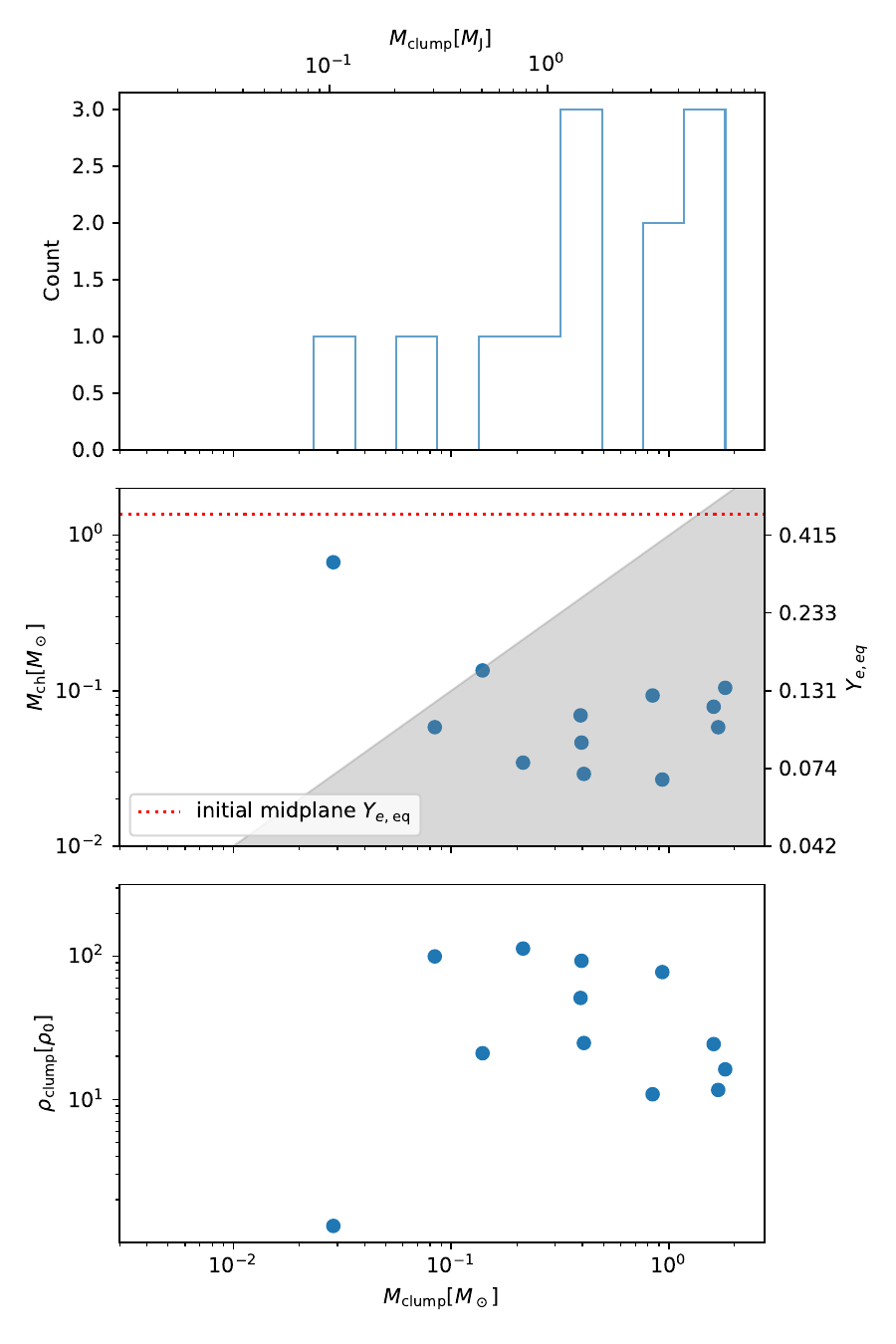}
\caption{Same as Fig.~\ref{fig:MchvsMclump}, but illustrating the bound clump properties for model \texttt{O8T1e10}, corresponding to an initially $\alpha$-rich disk.}
\label{fig:MchvsMclump_alpha}
\end{figure}

The statistics of the bound clumps for this simulation are shown in Figure \ref{fig:MchvsMclump_alpha}, 
following the same format as Fig.~\ref{fig:MchvsMclump}. 
The large reduction of the clump $Y_e$ below the disk's initial electron fraction $\approx 0.5$ (red dotted line) reduces the local Chandrasekhar mass, enabling the formation of NS of mass $\sim 0.1M_\odot$.  Other properties of the clump qualitatively resemble those in the higher $\dot{M}$ model \texttt{O16T1.5e10} (Sec.~\ref{sec:highdensity}). 
The final densities of the clumps reach $10$–$100 \rho_0$, 
with masses comparable to the local Jeans mass, 
though corresponding to smaller absolute masses given the different initial state ($\rho_0,T_0$) of the disk than in the higher $\dot{M}$ models.

\subsection{Summary Data}

\begin{figure*}[htbp]
\centering
\includegraphics[width=0.8\textwidth,clip=true]{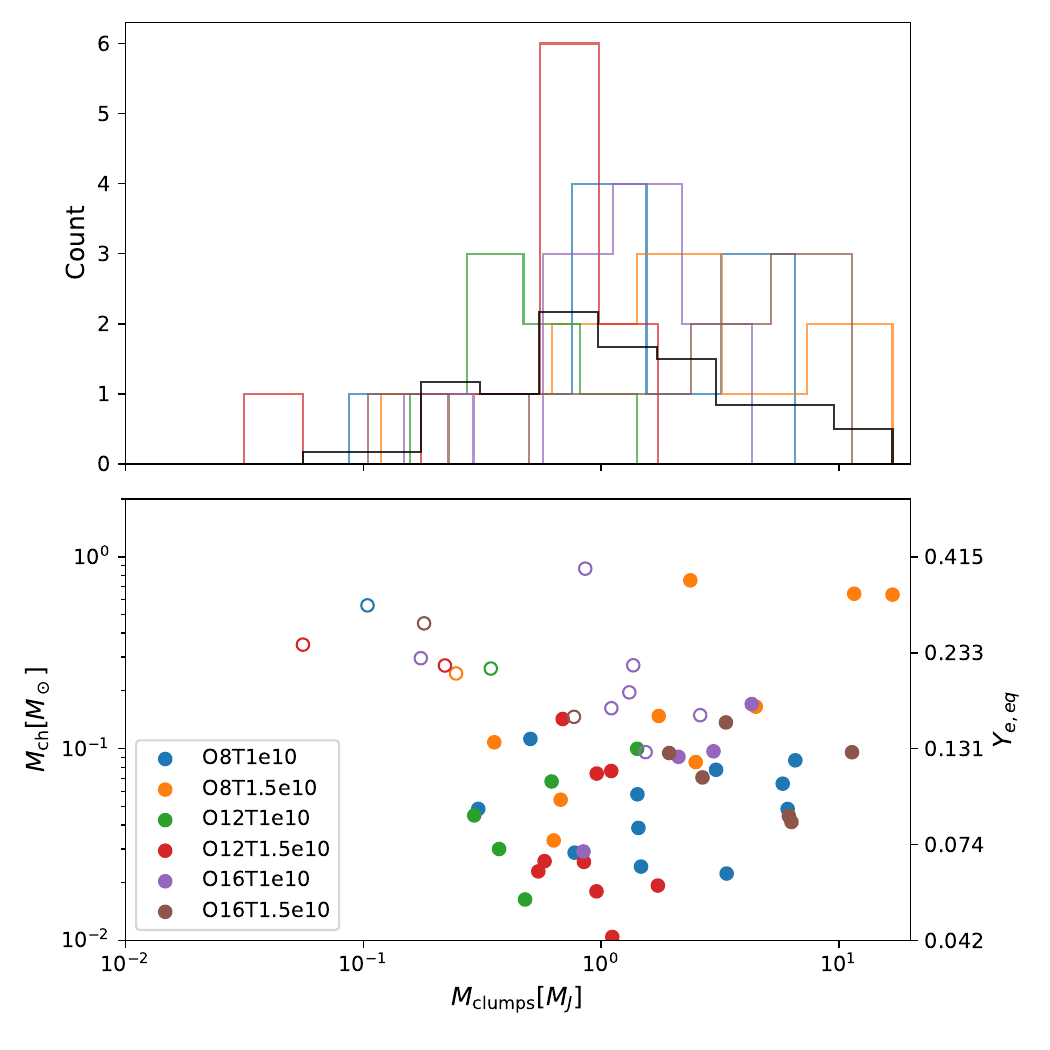}
\caption{Upper panel: Mass histogram (in units of the Jeans mass) of self-bound clumps similar to those shown in Figs.~\ref{fig:MchvsMclump}, \ref{fig:MchvsMclump_alpha} but for the broader suite of simulations. A black line shows the average of all 6 fragmenting simulations.  Lower panel: Average $Y_e$ and associated local Chandrasekhar mass $M_{\rm Ch}$ using the clump-averaged electron fraction (Eq.~\eqref{eq:MCh}). Filled circles indicate those clumps that exceed $M_{\rm Ch}$ and are therefore expected to continue collapsing to form NS. }
\label{fig:MchvsMclump_all}
\end{figure*}

Figure \ref{fig:MchvsMclump_all} summarizes the distribution of clump masses for our simulation suite. Most clumps are near or somewhat above the Jeans mass, but with a broad distribution that extends from $0.1-10M_{\rm J}$. 
Filled circles in the lower panel represent clumps with $M_{\rm clump} > M_{\rm ch}$, a condition that is more commonly satisfied by higher mass clumps. 
We conclude that the majority of the clumps across a relatively wide range of disk properties are likely to form NSs. 
Furthermore, 
because $Y_{e}$ becomes small within the high density clumps (regardless of the initial $Y_e$ of the disk midplane), $M_{\rm ch}$ 
is frequently small enough to allow sub-solar NSs to form.

In our fragmenting simulations, just a few clumps are sufficient to consume most of the mass within our simulation domain, leaving the residual gas density in the rest of the disk very low. This contrasts with the scenario proposed by \citet{Lerner2025}, 
who assume that over-densities form with a typical radii $\sim H/4$ (their Eq.~8), corresponding to an initial mass roughly 1/64 of the typical clump masses seen in our simulations. In their case, a few such objects do not deplete the disk significantly, allowing for continued migration driven by the disk, which will affect GW mergers that we discuss in \S \ref{sec:mergers}. This discrepancy raises concerns regarding the mass budget of realistic disks (see the $M_d(<r)$ column in Table \ref{tab:para}). Given that these very thick disks have $H \lesssim r$, 
the mass enclosed within the simulation domain ($\sim 36 \Sigma H^2$) can exceed $M_d(<r) = \pi \Sigma r^2$ by a factor of a few when scaled to a central black hole mass of $M_\bullet \sim 3M_\odot$ (though comparable for $M_\bullet  \sim 10M_\odot$). This means the mass of our simulated disks can surpass physical values for large disk radii $r$.  Imposing 
a realistic mass budget would reduce the number of clumps that can form, but even in that case, most of the mass would still end up in the bound objects rather than remaining in gas.

\section{Discussion}
\label{sec:discussion}

\subsection{Effects of alpha particle cooling}
\label{sec:alphacool}

Our simulations reveal that $\alpha$ particles are destroyed as clumps contract to high densities and temperatures (e.g., Fig.~\ref{fig:Xalphahighsigma}).  However, we have neglected the cooling associated with this dissociation, which occurs at a rate
\begin{equation}
\Lambda_{\alpha} \simeq -Q_{\alpha}(\rho/m_p)\frac{d}{dt}X_{\alpha}
\end{equation}
where $Q_{\alpha} \simeq 7.1$ MeV is the nuclear binding energy per baryon of an alpha particle.

\begin{figure}[htbp]
\centering
\includegraphics[width=0.42\textwidth,clip=true]{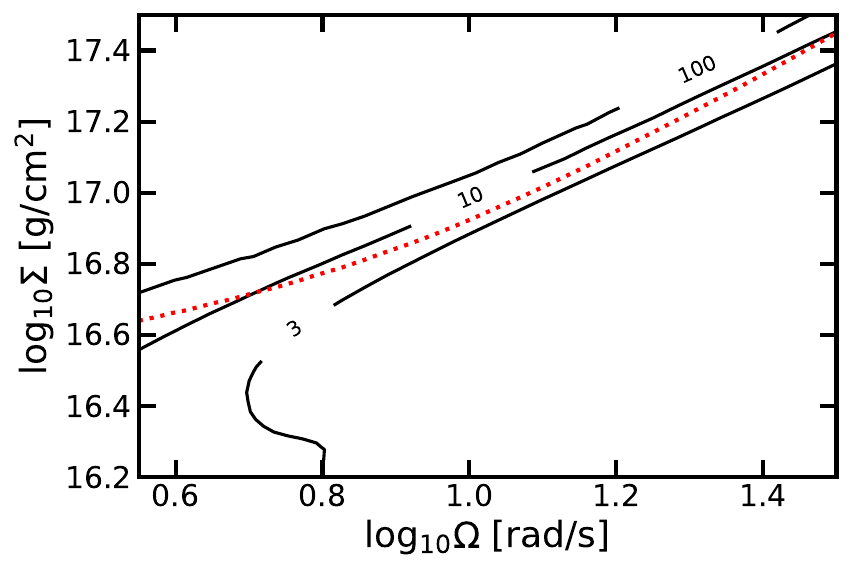}
\caption{Solid contours show the approximate dimensionless cooling timescale due to $\alpha$ particle dissociation, $\tau_{\rm cool,\alpha}$ (Sec.~\ref{sec:alphacool}), in the same parameter space of disk surface density and angular frequency shown in Fig.~\ref{fig:contours_tcool}.  We have very roughly estimated the $\alpha$ particle cooling rate for a contracting clump by assuming that $X_\alpha$ drops by an order-unity factor from its initial midplane value over one neutrino cooling timescale (see Eq.~\eqref{eq:dXalpha} and surrounding text). A comparison of these contours to those in Fig.~\ref{fig:contours_tcool} from neutrino cooling alone indicate the importance of $\alpha$ particle cooling, particularly for disks that begin rich in $\alpha$ particles ($X_{\alpha} \simeq 1$; below the dotted red line).}
\label{fig:taucool_alpha}
\end{figure}

We can roughly estimate the effect that alpha dissociation would have on our disk solutions as follows.  Analogous to the case of neutrino-cooling, we calculate an effective dimensionless cooling timescale due exclusively to alpha dissociation $\tau_{\rm cool,\alpha} \equiv U \Omega/\Lambda_\alpha$, under the assumption that $X_{\alpha}$ decreases by an order-unity fraction from its initial value over the first neutrino cooling timescale, i.e., 
\begin{equation}
\frac{1}{\Omega}\frac{\partial  X_\alpha}{\partial t} \sim - \dfrac{X_\alpha(t = 0)}{\tau_{\rm cool, 0}}.
\label{eq:dXalpha}
\end{equation}
Figure \ref{fig:taucool_alpha} shows $\tau_{\rm cool,\alpha} \equiv U \Omega/\Lambda_\alpha$ in the same $\Omega-\Sigma$ space shown in Fig.~\ref{fig:contours_tcool}.  For low-$\Sigma$ solutions, that begin alpha-rich in their initial states ($X_{\rm alpha} \simeq 1; \mathcal{G} < 1$), we see that $\tau_{\rm cool,\alpha}$ is even shorter than the neutrino cooling time alone ($\tau_{\rm cool}$ contours in Fig.~\ref{fig:contours_tcool}).  We conclude that if our simulations were to include the effects of alpha particle cooling, this would only expand the parameter space of disks undergoing runaway cooling and fragmentation. 

\subsection{Low Mass Neutron Star Mergers}
\label{sec:mergers}

Our calculations support the possibility that massive neutrino-cooled collapsar disks can fragment into bound objects spanning a wide range of masses $\sim 0.01-1 M_{\odot}$. 
Although our simulations cannot follow their complete dynamical collapse, 
this process will likely continue until finally being halted by neutron degeneracy pressure at densities $\gtrsim 10^{13}-10^{14}$ g cm$^{-3}$. 
Following a phase of diffusion-delayed (Kelvin-Helmholtz) cooling and contraction lasting seconds or less (e.g., \citealt{Pons+99}), these objects will form sub-solar NSs with radii $\sim 10-100$ km (Eq.~\eqref{eq:RNS}). 

Although our simulations cannot resolve this process either, 
it also seems likely that at least some of this swarm of disk-formed NSs will become paired into binaries (through e.g., gas dynamical friction; 
e.g., \citealt{DeLaurentiis+23,Dodici&Tremaine24}), possibly following a phase of radial migration driven by gas torques (e.g., \citealt{Piro&Pfahl07,Lerner2025}), although this possibility sensitively depends on how much gas remains in the disk after NS formation.
Over similar timescales of minutes, any formed NS binaries may merge together via a combination gas-driven migration and GW emission (\citetalias{metzger2024}). 
Collapsar-embedded NS mergers thus offer a potential source of sub-solar compact object potentially detectable by present and next-generation ground-based GW observatories \citep{Silva+16,Abbott_subsolar_22,Bandopadhyay+23}. 
The product of the merger of two low-mass NS is a NS with a gravitational mass slightly less than the sum of the original binary (e.g., \citealt{Giacomazzo+2013}). 
Although direct collisions or GW capture are inefficient given the small effective cross sections, 
circumbinary torques may significantly accelerate the coalescence once a binary is formed \citep{Li+2021,LiLai2022}, 
with gas dissipation enhancing the capture cross section \citep{Li+2023}. 
This mechanism supports the possibility of successive generations of NS mergers within the disk.

In Appendix \ref{sec:GWmerger} we estimate basic properties of the GW signal from subsolar NS mergers in the disk, in comparison to the simultaneous (and stronger) GW signal produced as the NS binary (or its merger product) coalesces with the central BH \citep{Piro&Pfahl07,Shahamat+21,Lerner2025}. 
Even in the pure vacuum point-mass approximation, sub-solar NS mergers are challenging to detect relative to ordinary NS mergers because the GW strain scales as the square of the masses of the merger bodies and because sub-solar NSs are physically larger $\gtrsim 30$ km, thereby causing the inspiral to terminate at lower frequency.  
Nevertheless, at distances of the nearest collapsars $\sim 100$ Mpc over a few year time window, 
binaries of mass $\gtrsim 0.1M_{\odot}$ are potentially detectable by LIGO/Virgo/Kagra and third-generation GW observatories.  

\begin{figure}[htbp]
\centering
\includegraphics[width=0.42\textwidth,clip=true]{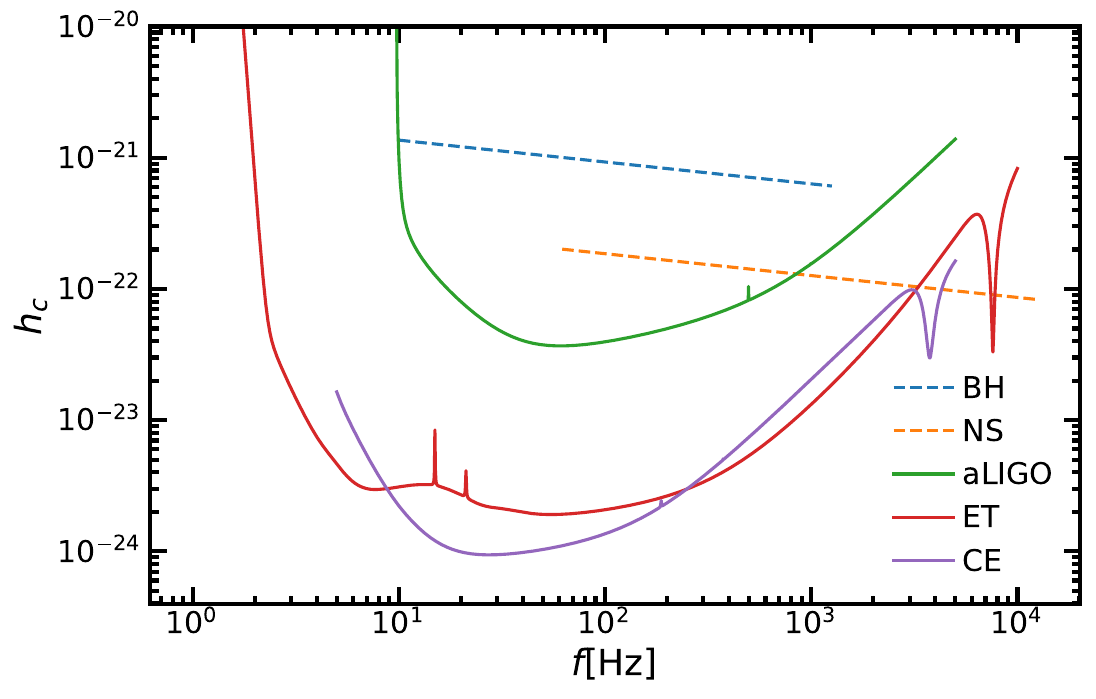}
\caption{Effective strain amplitude as a function of frequency $h_{\rm c}(f)$ (Eq.~\eqref{eqn:hc(f)}) for a sub-solar NS-NS (solid blue line) and subsequent NS-BH merger (solid orange line) in a collapsar disk at a distance of 100 Mpc.  We have assumed two NS of equal mass $M = 0.1M_{\odot}$ merging with a central BH of mass $M_{\bullet} = 3M_{\odot}$.  The NS binary is assumed to start from an initial separation equal to the Hill radius $r_{\rm H}$ (Eq.~\eqref{eq:rH}) and orbits at a distance $r = 200R_{\rm g}$ from the BH. For comparison we show the approximate $1\sigma$ sensitivity of current and third-generation ground-based GW observatories \citep{Iacovelli:2022bbs,Iacovelli:2022mbg}  If an appropriate signal template can be constructed, the square of signal-to-noise of an event is represented as the area between $h_c$ and detector noise on a log-$f$ plot.  For details see Appendix \ref{sec:GWmerger}.}
\label{fig:GW}
\end{figure}

The raw peak amplitude $h_{\rm NS} \gtrsim 10^{-25}$ is well below that of Advanced LIGO/Virgo but comparable to the projected strain sensitivity of Einstein Telescope \citep{Branchesi+23} and Cosmic Explorer \citep{Evans+21} at $\sim 10-300$ Hz. However, the detection prospects are improved substantially if matched filtering techniques can be applied. As shown in Figure \ref{fig:GW}, the effective strain sensitivity to the NS merger assuming GW-driven inspiral, $h_{c, \rm NS}$ (see Eqs.~\ref{eqn:h(t)}, \ref{eqn:f(t)}, \ref{eqn:hc(f)}; Appendix \ref{sec:GWmerger}), exceeds the raw strain sensitivity by a factor roughly equal to the square root of the number of orbital cycles spent in-band (e.g., \citealt{Flanagan&Hughes98}). Because $h_{c, \rm NS}$ exceeds the sensitivity curve of Advanced LIGO/Virgo by roughly a factor that decreases from $\sim 4$ to 0 over a width of $\Delta \ln f \sim 5$, 
the source would be marginally detectable with $S/N \approx 2.5$. 
The same event would achieve $S/N \sim 50-60 $ in Einstein Telescope/Cosmic Explorer.

However, the potential complexity of the waveforms could make matched filter techniques challenging to apply.
As long as NS binaries $\gtrsim 0.1 M_{\odot}$ start their GW-driven inspiral well within the Hill radius, at separations of a few NS radii, 
they will coalesce into a single body faster than the time it takes the binary itself to merge with the central BH. The NS merger chirp(s) will thus likely appear as short-lived high-frequency perturbation(s) around the more powerful but slowly evolving nearly period BH merger signal. 
The NS merger chirp will also experience Doppler modulation (with an amplitude of around $\sim 10\%)$ due to the orbital motion of the NS binary around the BH, as shown in Figure \ref{fig:GWwaveform}. Several additional effects on GW waveform, such as higher-order post Newtonian corrections, tidal effects between the NS or between the NS binary and the BH, Doppler modulation of the strain amplitude, binary eccentricity, gas effects and Shapiro time-delays, all of which can further complicate the waveforms and prevent simple construction of matched filter templates.

\subsection{Very Low Mass (``Forbidden'') Neutron Stars}
\label{sec:minmass}
Our simulations find fragments with masses much smaller than the usually considered minimum dynamically stable NS mass of $M_{\rm min} \approx 0.1M_{\odot}$ (e.g., \citealt{Haensel+02}).  However, the stability analysis leading to this minimum generally assumes $\beta-$equilibrium and hence may not strictly apply in the collapsar environment, where the system can in principle evolve on timescales faster than $\beta$-decays.  This opens the possibility for the existence of NSs with ``forbidden'' masses $< M_{\rm min}$, whose gravitational wave signal could offer evidence for formation in a collapsar environment.

Over longer timescales after formation, the growing degeneracy pressure enabled by $\beta$-decays will cause such sub-$M_{\rm min}$ NS to expand, potentially leading to a dynamical explosion \citep{Page82,Colpi+89,Colpi&Rasio94}.  However, the timescale for this expansion to occur is set by the $\beta-$decay rate of nuclei present in the star (\citealt{Colpi+89}).  The latter was estimated by \citet{Colpi+89} to range from milliseconds to several seconds given the nuclei present for a cold equations of state; however, the latter may not be justified given the high temperatures in a newly-formed proto-NS and hence could in principle be as long as the free neutron decay timescale of 10 minutes.  

The inflation or explosion of very low-mass NS inside collapsar disks, could lead to novel observational consequences. For example, the eventual Roche lobe overflow of these bodies onto the central black hole could lead to their tidal disruption and accretion, causing punctuated episodes of accretion that induce variability in the GRB jet (e.g., \citealt{Perna+06}).  Future studies of the long-term evolution of $< M_{\rm min}$ neutron stars, endowed with hot initial conditions specific to their formation in collapsar disks, are necessary to assess their lifetimes and potential fates.

\section{Conclusions}
\label{sec:conclusion}

We have presented three-dimensional shearing-box hydrodynamical simulations of self-gravitating neutrino-cooled accretion disks, applicable to the outer regions of collapsar disks.  We find that for marginally stable initial disks ($Q_{T} = 1),$
runaway collapse occurs for dimensional cooling timescale $\lesssim 10$.  Disks with these properties are achieved for infall/accretion rates $\gtrsim M_\odot$ s$^{-1}$, in the range expected in the earliest stages following the core-collapse of very massive rotating stars.

Our fragmenting models generate clumps with initial masses on the order of the local Jeans mass, 
with the total surface density of the clumps becoming larger and their mass spectrum wider for shorter cooling times. Even if the disk itself is not neutron-rich in the initial state, contracting clumps obtain low $Y_e \ll 0.5$ from electron capture reactions on protons; as a result, most clumps outweigh their local Chandrasekhar mass by the end of the simulation and hence are likely to go on to form NS with masses well below solar. The free-fall time of the clumps is comparable to the disk's dynamical time of seconds or less, similar to the Kelvin-Helmholtz cooling time over which any newly formed NSs will settle into their final cold radii. The standard wisdom that stable NSs do not exist with masses below $\lesssim 0.1M_{\odot}$ may not apply in the collapsar environment on timescales faster than $\beta-$ decay timescale, enabling the possibility of transiently stable ``forbidden'' very-low mass NS (Sec.~\ref{sec:minmass}).

When reporting our results in physical units, e.g. for the disk radius or Jeans mass, we have defaulted to black hole mass of $3M_{\odot}$.  However, most of our results, such as the portion of $\Sigma-\Omega$ parameter space that results in fragmentation (Fig.~\ref{fig:contours_tcool}), are not sensitive to this assumption and also apply to more massive black holes, even those $\gtrsim 100 M_{\odot}$ comparable to the largest black holes detected by LIGO (e.g., \citealt{LIGO2025}).  If such spinning compact objects are formed from the collapse of very massive rotating stars (``super-collapsars''; \citealt{Siegel+22}), the threshold for instability is generally achieved closer to the black hole horizon (in gravitational radii) and the fragmentation is expected to form greater mass bodies ($M_{\rm J} \propto M_{\bullet}$; Eq.~\eqref{eq:MJ}). Given also the higher mass-infall rates achieved by the collapse of very high-mass stars, fragmentation is likely easier to achieve than for lower-mass collapsars, even while such events are probably much rarer in the universe.

Although our simulations do not address this possibility directly, we hypothesize that at least some of the disk-formed NSs will pair into bound binaries, ultimately merging and emitting GWs in frequency ranges accessible to ground based gravitational wave observatories. These chirp signals could in principle be picked up by LIGO/Virgo's subsolar mass compact object coalescence searches for particularly nearby events (e.g., \citealt{Abbott_subsolar+18,Abbott_subsolar_22}). Although sub-solar mergers could also arise from the coalescence of primordial BHs, the strong tidal effects of low-mass NS mergers \citep{Silva+16,Bandopadhyay+23,Crescimbeni+24}, as well as effects on the GW waveform due to the orbital motion of the binary around the central BH such as Doppler modulation (\citetalias{metzger2024}; Fig.~\ref{fig:GW}), can help distinguish these possibilities. The disk itself may produce other GW signals due to Rossby waves \citep{Gottlieb+24} or spiral density waves associated with non-fragmenting gravitationally unstable portions of the disk \citep{Siegel+22}.
However, given the rarity of collapsars in the local universe and the resulting large distances to the closest such events $\gtrsim 100$ Mpc, the first detections as GW sources might requiring waiting for the greater sensitivity of third-generation observatories such as Einstein Telescope or Cosmic Explorer.  

NS mergers that occur in the collapsar disk will give rise to neutron-rich ejecta (\citetalias{metzger2024}), a portion of which could achieve sufficiently high velocities to add to any neutron-rich outflows that occur from disk winds or jets \citep{Siegel+19,Issa+25}, impacting the opacity of the inner supernova ejecta and reddening its observed colors similar to a kilonova \citep{Barnes&Metzger22}.  The rough temporal coincidence of a hierarchical chain of sub-solar NS mergers with the long GRB and supernova explosion powered by the collapsar disk, would provide a ``multi-messenger symphony'' unique in the cosmos (\citetalias{metzger2024}).  The ``grand finale'' to the symphony occurs as the resulting merger product(s) gradually coalesce with the central BH in a final loud chirp signal(s), the scenario first described by \citet{Piro&Pfahl07}.  

To further investigate these processes, our framework can be extended to global simulations that include cooling driven from $\alpha$ particle dissociation and heating from nuclear burning \citep{Zenati+20}. Adaptive mesh refinement and/or sub-grid sink particle methods could help properly resolve the clumps, 
overcome fragmentation bottlenecks caused by extremely short numerical timesteps, and enable following the pairing and mergers of the collapsed NS.  Separate studies should explore disk-driven migration of the single NS or merger products towards the central BH \citep{Piro&Pfahl07}, as recently explored semi-analytically by \citet{Lerner2025}.

We thank the reviewer for providing insightful comments, which improved the presentation of the manuscript. Y.X.C would like to thank Jeremy Goodman, Matteo Cantiello, Douglas Lin and Zhiwei Chen for helpful discussions. B.~D.~M. acknowledges partial support from the National Science Foundation (grant number AST-2406637) and the Simons Foundation (grant number 727700).  The Flatiron Institute is supported by the Simons Foundation. 
We acknowledge computational resources provided by the high-performance computer center at Princeton University, which is jointly supported by the Princeton Institute for Computational Science and Engineering (PICSciE) and the Princeton University Office of Information Technology.

%




\appendix
\section{Equation of State and Neutrino Cooling Rate}
\label{app:EOS}

Here we describe the equation of state (EOS) and the optically-thin neutrino cooling rate used in our simulations.  The assumed composition includes photons, electrons, positrons, free nucleons (neutrons and protons), and $\alpha$ particles in nuclear statistical equilibrium.  The disk is assumed to be locally optically-thin to neutrinos, as is generally a good approximation in those regions of the disk where self-gravity is important \citep{ChenBel2007, Liu2017, Lerner2025}, so the neutrino chemical potentials are assumed to be zero. In general the EOS and cooling rates depend on the electron fraction $Y_e$ in addition to density and temperature; however, we shall assume that the $e^{\pm}$ captures that dominate the weak interactions in the disk are sufficiently rapid to hold $Y_e$ close to an equilibrium value $Y_{e,eq}(\rho,T)$, which we calculate as described below.  With these assumptions, both the thermodynamic quantities entering the EOS and the cooling rates become functions of $\rho$ and $T$ alone.

The total number densities of protons and neutrons, respectively, including those contained within $\alpha$ particles, are given by
\begin{equation}
    n_p = Y_e \dfrac{\rho}{m_p}, n_n = (1-Y_e) \dfrac{\rho}{m_p}.
\end{equation}
In addition to the electrons that accompany the protons, electron/positron pairs are produced copiously through the reaction $e^{-} + e^{+} \leftrightarrow \gamma + \gamma$.  The number densities of electrons and positrons are given by
\begin{equation}
    n_{e^{ \pm}}=\frac{\left(m_e c\right)^3}{\pi^2 \hbar^3} \int_0^{+\infty} f\left(\sqrt{p^2+1}, \mp \mu \right) p^2 d p,
\end{equation}
where
\begin{equation}
f(E, \mu)=\frac{1}{e^{\frac{E-\mu}{\theta}}+1}, \theta(T) = kT/m_e c^2 
\end{equation}
is the Fermi-Dirac distribution for an assumed electron chemical potential $\mu$. The value of $\mu$ is determined by the charge neutrality condition,
\begin{equation}
    n_{e^{-}}(\mu, T) -n_{e^{+}}(\mu, T)=Y_e \frac{\rho}{m_p}.
\end{equation}

As mentioned above, we assume that $Y_e$ reaches an equilibrium value $Y_{e,eq} = Y_e(T, \rho)$ determined by the balance between electron captures on free protons ($e^{-} + p \rightarrow \nu_e + n$) and positron captures on free neutrons ($e^{+} + n \rightarrow \bar{\nu}_e + p$). In other words, $Y_e = Y_{e,eq}$ is set by condition
\begin{equation}
\dot{n}_{e^{+} }(\rho, T, \mu, Y_e)=\dot{n}_{e^{-} }(\rho, T, \mu, Y_e),
\end{equation}
where the $e^{\pm}$ pair capture rates are given by
\citep{Beloborodov2003,ChenBel2007}
\begin{equation}
\dot{n}_{e^-} = 
(n_p - 2n_\alpha) K \int_0^{+\infty} f\left(E+Q_{np}, \mu\right)(E+Q_{np})^2\left[1-\frac{1}{(E+Q_{np})^2}\right]^{1 / 2} E^2 d E;
\label{eqn:EOSneminus}
\end{equation}
\begin{equation}
\dot{n}_{e^+}= (n_n - 2n_\alpha) K \int_{Q_{np}+1}^{+\infty} f\left(E-Q_{np}, -\mu\right)(E-Q_{np})^2\left[1-\frac{1}{(E-Q_{np})^2}\right]^{1 / 2} E^2 d E,
\label{eqn:EOSneplus}
\end{equation}
$Q_{np}=\left(m_n-m_p\right) / m_e=2.53$ is the neutron-proton mass difference and $K=6.5 \times 10^{-4} \mathrm{~s}^{-1}$ is a constant related to the measured decay rate of the neutron. Note from the prefactors that the $e^{\pm}$ pair capture rates depend only on the densities of {\it free} protons and neutrons because $\alpha$ particles do not capture pairs.

The $\alpha$ particle density is given by \begin{equation}
    n_\alpha = \rho\dfrac{X_\alpha}{4m_p} ,
\end{equation}
where the alpha mass fraction $X_\alpha (Y_e, T, \rho)$ follows from NSE \citep{Meyer1994}
\begin{equation}
4.9 \times 10^2 \rho_{10}^{-3 / 2} T_{10}^{9 / 4} \exp \left(-\frac{16.4}{T_{10}}\right)=4\left[Y_e-\frac{X_\alpha}{2}\right]\left[1-Y_e-\frac{X_\alpha}{2}\right]X_\alpha^{-1 / 2}, \rho_{10} = \dfrac{\rho}{10^{10 }{\rm g/cm}^3}, T_{10} = \dfrac{T}{10^{10} {\rm K}}.
\end{equation}
The maximum $\alpha$ mass fraction for a given electron fraction $Y_e$ is $X_{\alpha} = 2Y_{e}$.

\begin{figure*}[htbp]
\centering
\includegraphics[width=0.32\textwidth,clip=true]{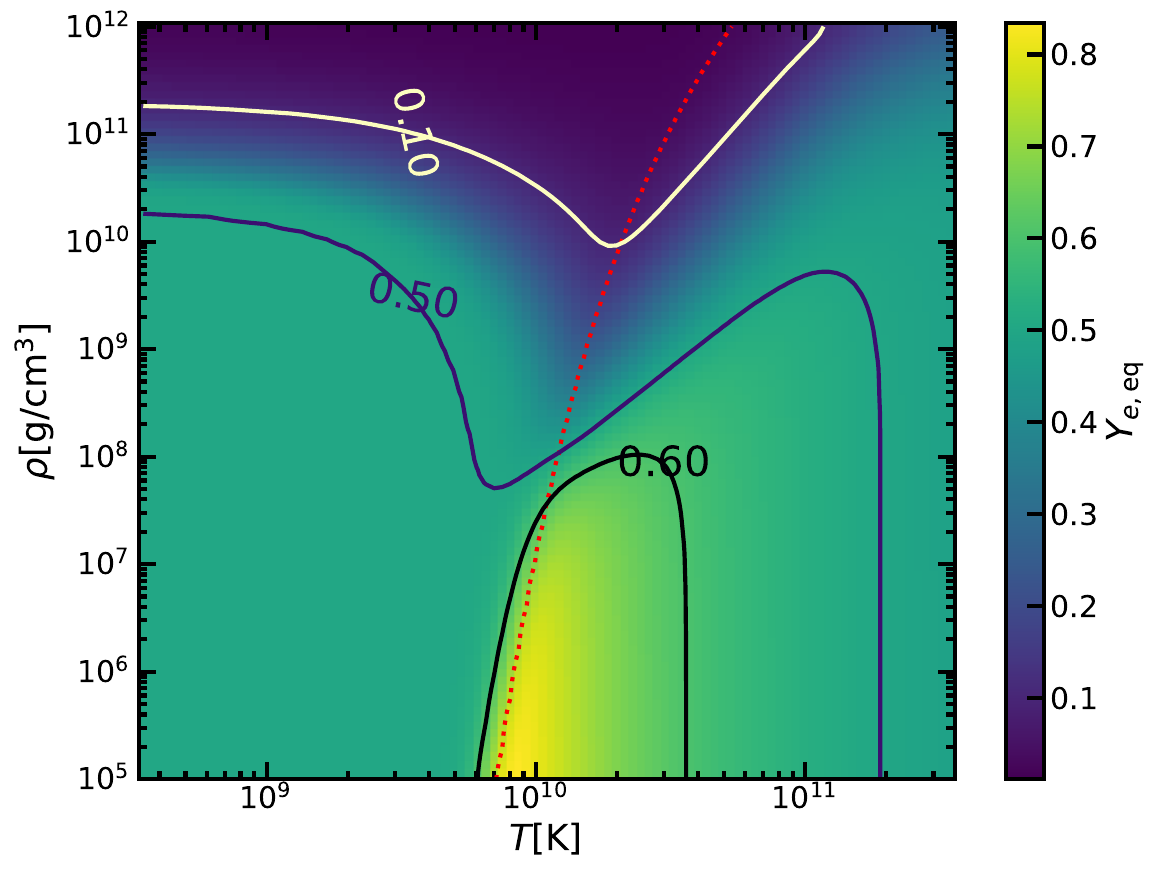}
\includegraphics[width=0.32\textwidth,clip=true]{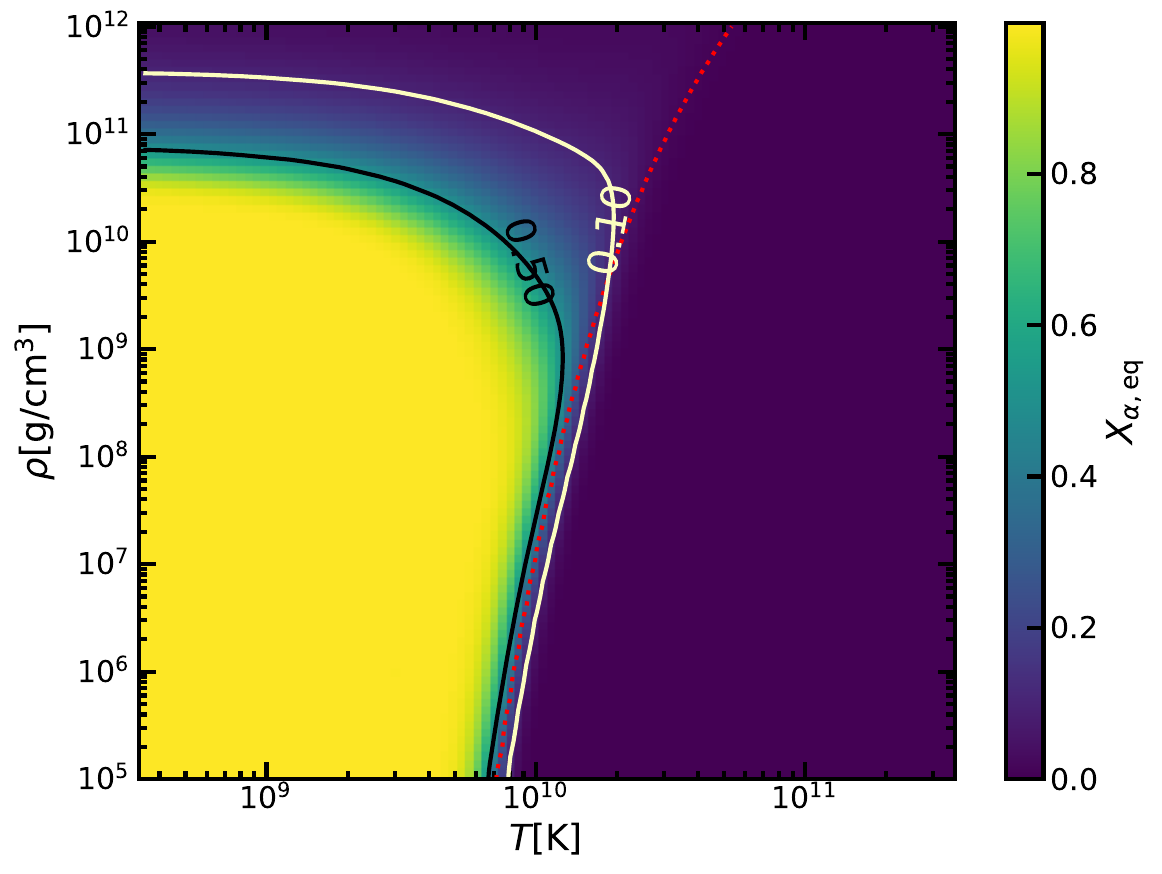}
\includegraphics[width=0.32\textwidth,clip=true]{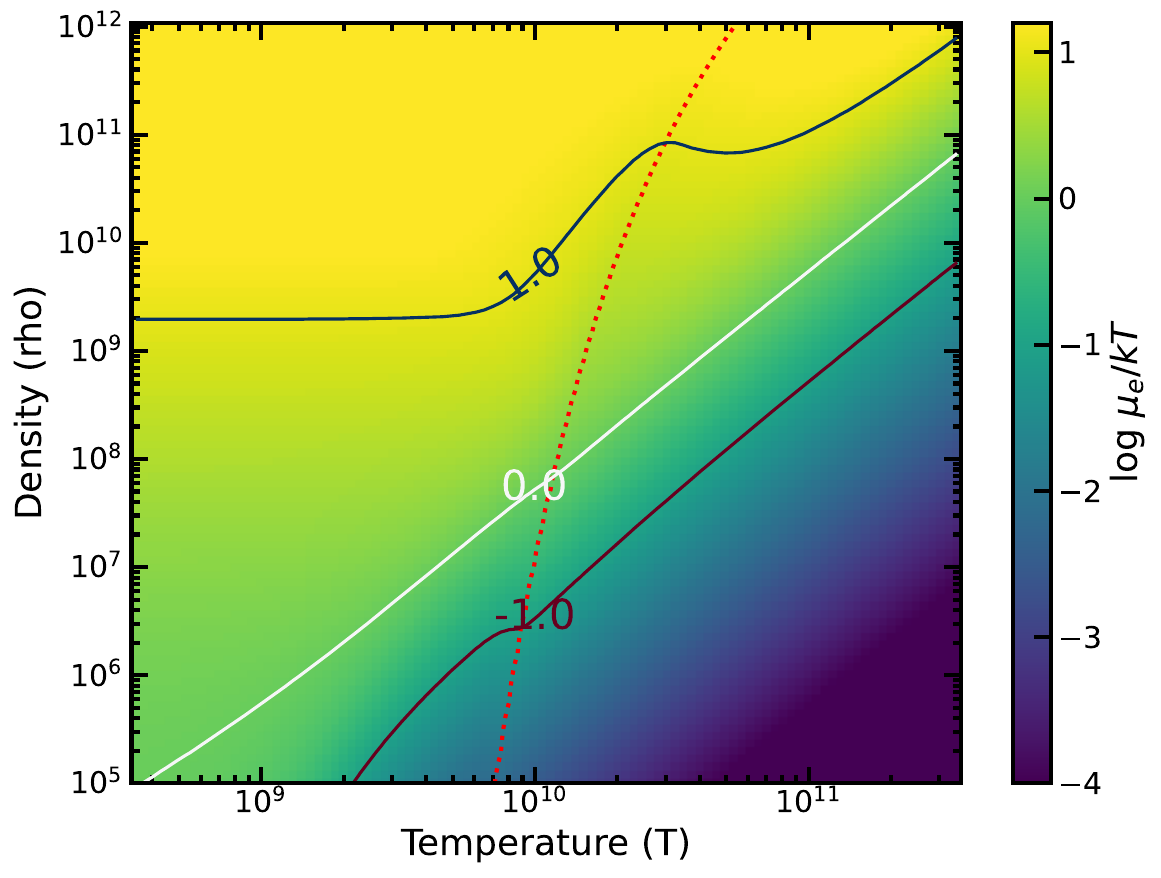}
\caption{Equilibrium electron fraction $Y_{e,\rm eq}$, $\alpha$ particle mass fraction $X_{\alpha}$, and electron chemical potential $\mu/kT$ in the space of $\rho$ and $T$.  A red dotted line shows the contour $\mathcal{G} = 1$ (Eq.~\eqref{eqn:GrhoT}) separating the regions of parameter space where all nucleons are free from that where most protons are trapped in $\alpha$ particles.}
\label{fig:Yeeq-rho-T}
\end{figure*}

The left panel of Figure \ref{fig:Yeeq-rho-T} shows contours of $Y_{e, \rm eq}(\rho,T)$ in the space of density and temperature, while the right panel shows contours of $X_{\alpha}(\rho,T)$. An upper limit on the $\alpha$ particle mass fraction $X_{\rm \alpha, \rm max}$ occurs at $Y_e = 0.5$ for the combination of $\rho, T$ satisfying 
\begin{equation}
    \mathcal{G}(\rho, T) = 4.9 \times 10^2 \rho_{10}^{-3 / 2} T_{10}^{9 / 4} \exp \left(-\frac{16.4}{T_{10}}\right) =  \dfrac{(1-X_{\alpha,\rm max})^2 }{\sqrt{X_{\alpha,\rm max}}}
    \label{eqn:GrhoT}.
\end{equation}

The contour $\mathcal{G} = 1$ denoted by a red dotted line in Figure \ref{fig:Yeeq-rho-T} delineates those parts of parameter space where alpha particle formation is relevant to the disk composition.
For $\mathcal{G} \gg 1$ (right of this red dotted line), we have $X_{\alpha} \ll 1$ and the $Y_{e, \rm eq}$ contours agree well with \citet[their Fig.~1]{Beloborodov2003}.  On the other hand, to the left of the red line where, $\mathcal{G} \ll 1$, the $\alpha$ particle mass fraction is close to its maximum value $X_{\alpha} \simeq 2Y_e$.  At low densities and temperatures where $Y_{e} \simeq 0.5$ we see that $X_{\alpha} \simeq 1$, i.e. all the mass is locked into $\alpha$ particles. In contrast, at high densities and low temperatures, although the alpha mass fraction still reaches its maximum value $X_{\alpha} = 2Y_{e}$ give the number of protons available, this value is small because $Y_{\rm e,eq} \ll 0.5$ (the high electron degeneracy $\mu/kT \gg 1$ at high $\rho$ favors electron captures over positron captures). This has the implication that a bound clump undergoing runaway cooling in the disk to high densities will see both $Y_{\rm e,eq}$ and $X_{\alpha}$ drop in tandem, even if  $\mathcal{G} \ll 1$ (for an example, see the simulation \texttt{O8T1e10} shown in Fig.~\ref{fig:Xalphahighsigma}).

The EOS is determined by the total pressure, 
\begin{equation}
P = P_b + P_r + P_{e^+} + P_{e^-},  
\end{equation}
and internal energy, 
\begin{equation}
U = U_b + U_r + U_{e^+} + U_{e^-}, 
\end{equation}
which receive contributions from baryons,
\begin{equation}
P_b=\frac{\rho}{m_p} k_B T\left(1-X_\alpha+\frac{X_\alpha}{4}\right), \quad U_b=\frac{3}{2} P_b,
\end{equation}
radiation 
\begin{equation}
    P_r = a_r T^3/3, U_r = 3P_r,
\end{equation}
(where $
a_r=7.56 \times 10^{-15} \mathrm{erg} \mathrm{~cm}^{-3} \mathrm{~K}^{-4}
$), electrons, and positrons.  Using $\mu(\rho,T)$ and $Y_e = Y_{e,eq}(\rho,T)$, we calculate the pressure and internal energy contributions of the electrons and positrons according to:
\begin{equation}
\begin{aligned}
&P_{e^{ \pm}}=\frac{1}{3} \frac{\left(m_e c\right)^3}{\pi^2 \hbar^3} m_e c^2 \int_0^{+\infty} f\left(\sqrt{p^2+1}, \mp \mu\right) \frac{p^4}{\sqrt{p^2+1}} d p\\
&U_{e^{ \pm}}=\frac{\left(m_e c\right)^3}{\pi^2 \hbar^3} m_e c^2 \int_0^{+\infty} f\left(\sqrt{p^2+1}, \mp \mu \right) (\sqrt{p^2+1}-1) p^2 d p
\end{aligned}
\end{equation}

\begin{figure*}[htbp]
\centering
\includegraphics[width=0.32\textwidth,clip=true]{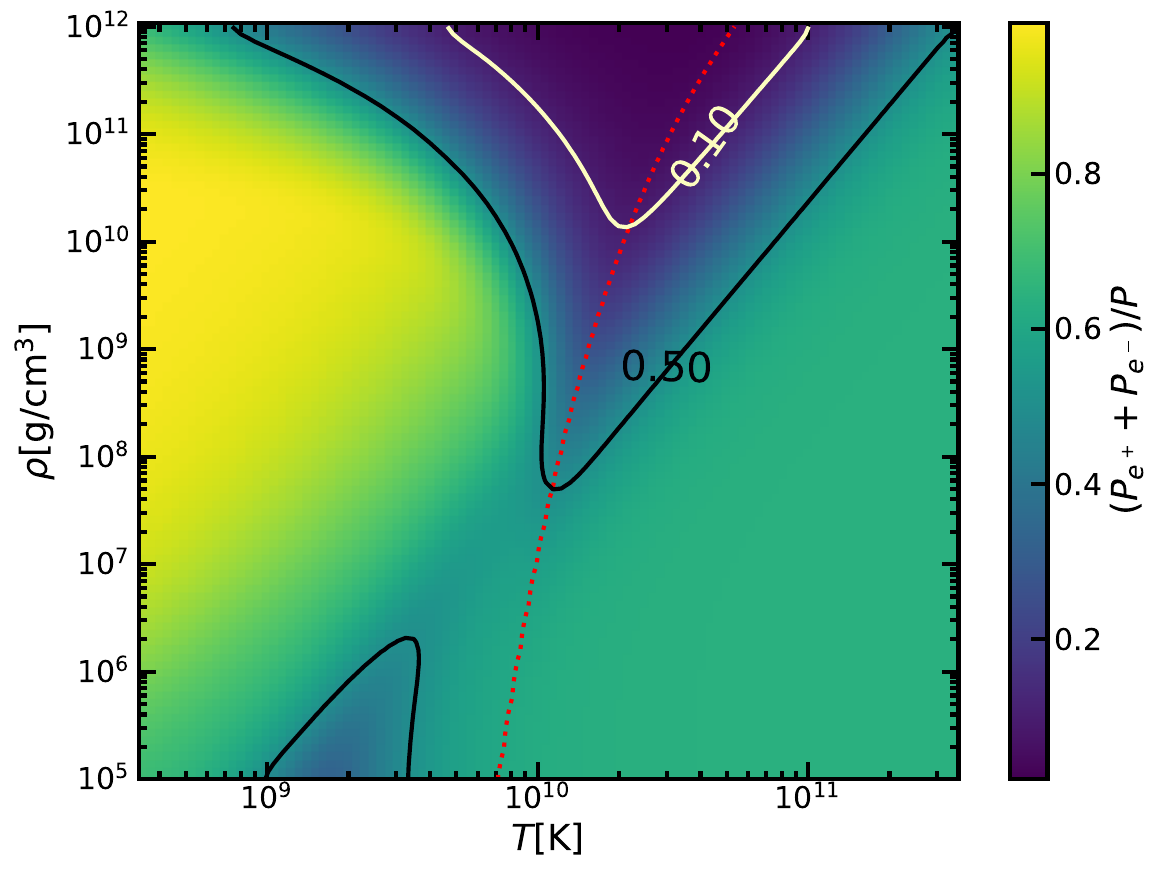}
\includegraphics[width=0.32\textwidth,clip=true]{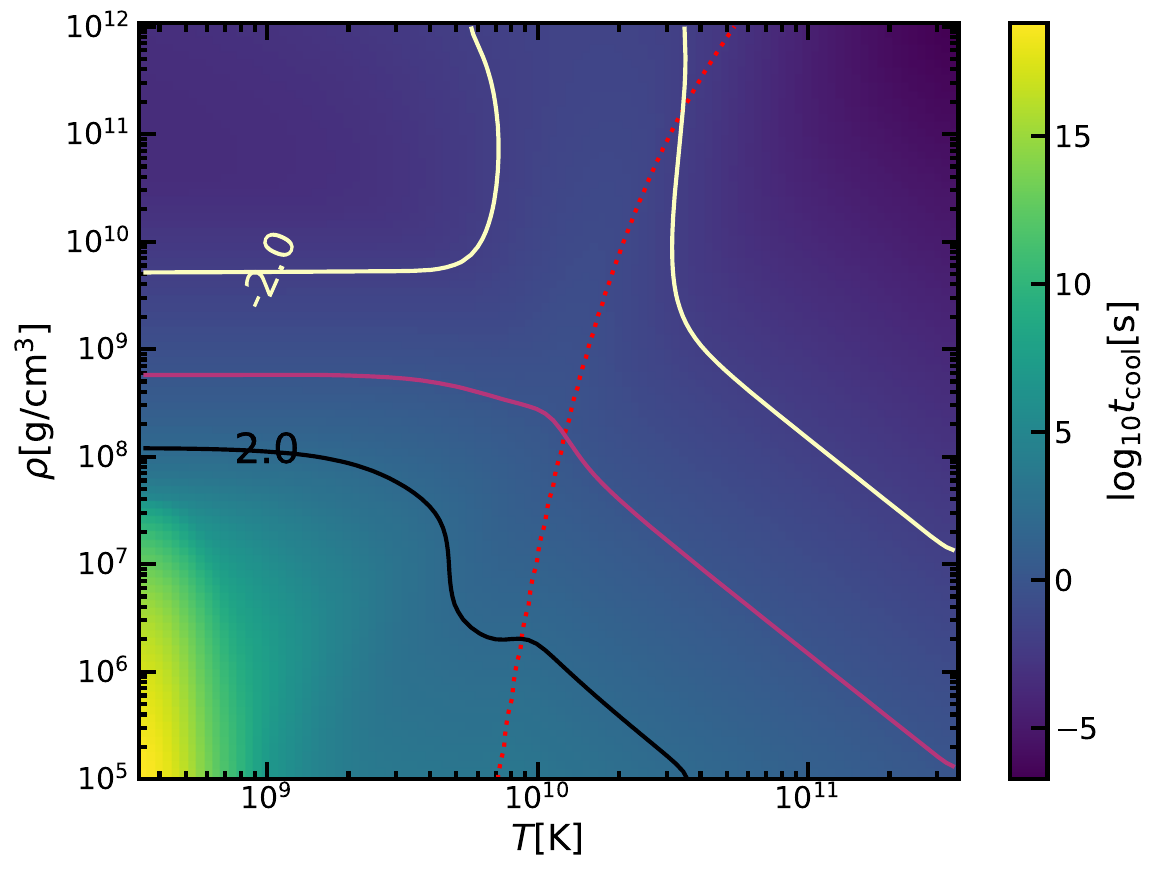}
\includegraphics[width=0.32\textwidth,clip=true]{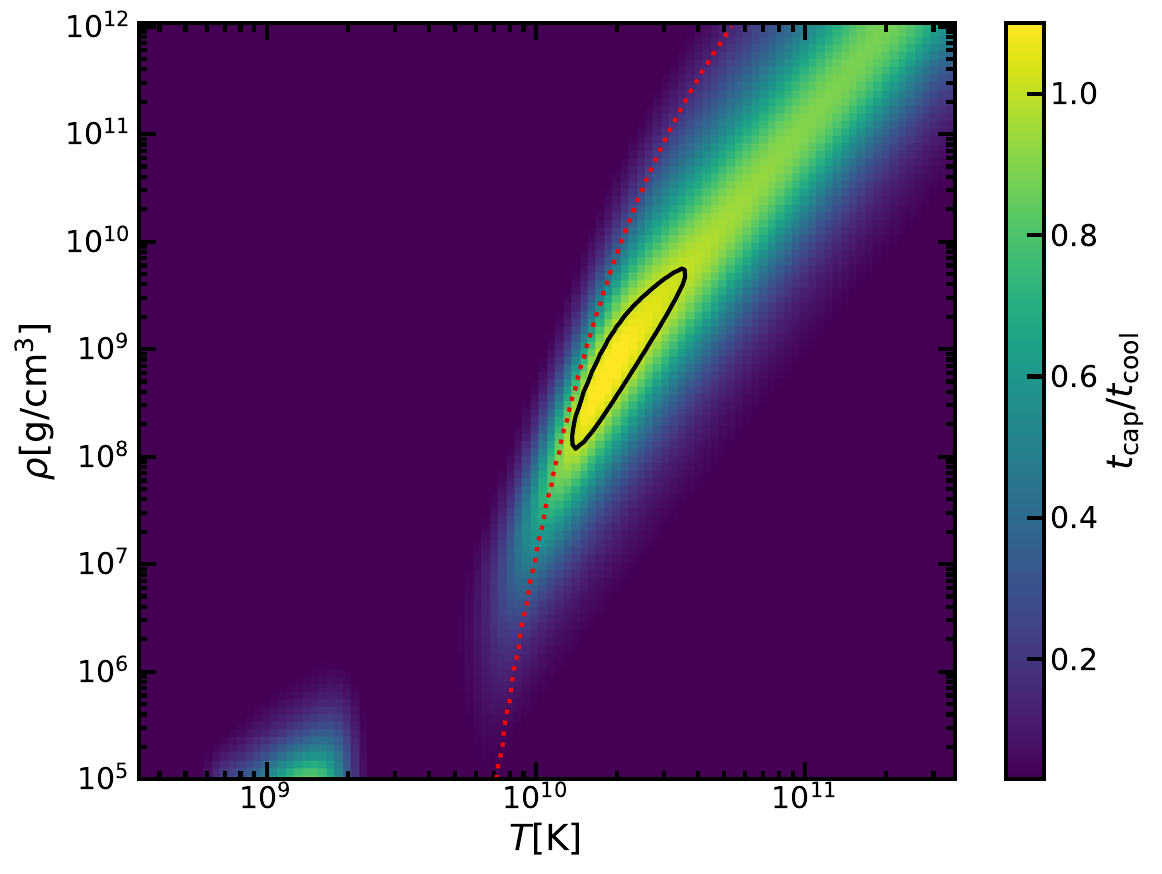}
\caption{Electron/positron pressure fraction (left panel), thermal cooling time $t_{\rm cool}$ in seconds (middle panel) and in ratio to the electron capture timescale $t_{\rm capt}$ (Eq.~\eqref{eq:tcapt}; right panel) in the $\rho-T$ space.  The fact that $t_{\rm capt} \ll t_{\rm cool}$ throughout the majority of the parameter space suggests that electron capture reactions are typically fast compared to the cooling rate of the plasma, justifying our assumption of an equilibrium electron fraction in the disks of interest where neutrino cooling is relevant. A red dotted line again denotes the contour $\mathcal{G} = 1$ (Eq.~\eqref{eqn:GrhoT}) corresponding to significant $\alpha$ particle formation.}
\label{fig:tcool}
\end{figure*}

The left panel of Figure \ref{fig:tcool} shows the fraction of pressure supplied by electrons and positrons relative to the total pressure. In the low $Y_{e, \rm eq}$ region where fragments typically lie, the degeneracy pressure is subdominant.

The total neutrino cooling rate $\Lambda = {\Lambda}_{e^-} + {\Lambda}_{e^+}$ is likewise given by the sum of the contributions from the same $e^{\pm}$ captures on free nucleons whose balance determined $Y_e$: 

\begin{equation}
{\Lambda}_{e^-}= (n_p - 2 n_\alpha)
m_e c^2 K \int_0^{+\infty} f\left(E+Q_{np}, \mu\right)(E+Q_{np})^2\left[1-\frac{1}{(E+Q_{np})^2}\right]^{1 / 2} E^3 d E
\end{equation}

\begin{equation}
{\Lambda}_{e^+}= (n_p - 2 n_\alpha) m_e c^2  K \int_{Q_{np}+1}^{+\infty} f\left(E-Q_{np}, -\mu\right)(E-Q_{np})^2\left[1-\frac{1}{(E-Q_{np})^2}\right]^{1 / 2} E^3 d E
\end{equation}

The middle panel of Figure \ref{fig:tcool} shows the cooling timescale, $t_{\rm cool} = U/\Lambda$ in the space of ($\rho,T$). At high temperatures the cooling rate scales sensitively with the temperature $\Lambda\propto T^6$ \citep{Beloborodov2003} and hence the cooling time becomes very short. Likewise, at low temperatures and high densities where electrons are relativistic and degenerate, $\Lambda \propto \rho$, and $t_{\rm cool}$ again becomes short. One can also define a characteristic electron capture timescale,
\begin{equation}
    t_{\rm cap} = \dfrac{n_p-2n_\alpha}{\dot{n}_{e^-}},
    \label{eq:tcapt}
\end{equation}
over which $Y_e$ will reach its equilibrium value starting from a higher initial value (e.g., $Y_e = 0.5$ for the infalling star in a collapsar). The right panel of Figure \ref{fig:tcool} shows the ratio $t_{\rm cap}/t_{\rm cool}$, which we see is $\ll 1$ across most of the $\rho-T$ space, apart from a narrow region outlined by a black contour. For the neutrino-cooled disks of interest, $t_{\rm cool}$ is also short compared to the radial gas inflow time through the disk (e.g., \citealt{ChenBel2007}).  The fact that $t_{\rm cap} \ll t_{\rm cool}$ provides a rough justification for our assumption that $Y_e \approx Y_{e,eq}$ everywhere in the disk (see also \citealt{Beloborodov2003}). 

Finally, our simulations neglect cooling associated with the dissociation of $\alpha$ particles \citep{Piro&Pfahl07}.  This can become relevant as disk fragments undergo runaway collapse to high densities, particularly in cases where the disk material begins alpha-rich in its initial state. The cooling from alpha dissociation will only further accelerate the collapse due to neutrino cooling alone (Fig.~\ref{fig:taucool_alpha}), thus further expanding the parameter space of fragmenting disk solutions (see Sec.~\ref{sec:alphacool} for further discussion). 

\section{Gravitational Waves from Mergers in Collapsars}
\label{sec:GWmerger}

Neglecting general relativistic corrections, the radius of a cold NS of mass $M \ll M_{\odot}$ and uniform electron fraction $Y_e$ is approximately given by (e.g., \citealt{Shapiro&Teukolsky83})
\begin{equation}
R_{\rm NS} \simeq 33\,{\rm km}\left(1-Y_{e}\right)^{-1/3}\left(\frac{M}{0.1M_{\odot}}\right)^{-1/3} .
\label{eq:RNS}
\end{equation}
This estimate assumes a NS supported entirely by non-relativistic neutron degeneracy pressure, a good approximation given the low central density $\ll 10^{13}$ g cm$^{-3}$ for sub-solar mass NS.  

We imagine that two such NS meet each other in the disk and are bound and driven close together through a combination of processes, possibly including gas-induced migration.  Consider the simplest case of a circular binary of two equal mass NS, which itself orbits the central BH of mass $M_{\bullet}$ at a radius $r \sim 100R_{\rm g}$ characteristic of our fragmenting disk solutions (Fig.~\ref{fig:contours_tcool}).  The NS binary will merge through through the emission of GWs on a timescale of \citep{Peters64}

\begin{equation}
\tau_{\rm NS} = \frac{5}{256}\frac{c^{5}a^{4}}{G^{3}M^{3}} \simeq 2\times 10^{3}\,{\rm s}\,\left(\frac{a_0}{100\,{\rm km}}\right)^{4}\left(\frac{M}{0.1M_{\odot}}\right)^{-3},\,
\end{equation}

where $M$ is the mass of each NS and $a_0$ is the initial binary semi-major axis at which GWs take over the system evolution.  The latter cannot exceed the Hill radius,
\begin{equation}
R_{\rm H} \simeq r\left(\frac{2M}{3M_{\bullet}}\right)^{1/3} \simeq 250\,{\rm km}\,\left(\frac{r}{200R_{\rm g}}\right)\left(\frac{M_{\bullet}}{3M_{\odot}}\right)^{2/3}\left(\frac{M}{0.1M_{\odot}}\right)^{1/3},
\label{eq:rH}
\end{equation}

However, $a_0 \ll R_{\rm H}$ is possible if the gaseous environment of the collapsar disk drives the binary together, e.g. through migration torques (similar to EMRIS or BH binaries in AGN disks; e.g., \citealt{Yunes+11,Stone+17,Bartos+17, Lerner2025}). However we caution there might not be much gas left to speed up the mergers if most of the disk's mass is consumed in forming the bound fragments.

The GW signal will reach its maximum just prior to the merger, as occurs at a binary separation $a_{\rm f} \simeq 2 R_{\rm NS}$ (Eq.~\eqref{eq:RNS}), at an approximate frequency

\begin{equation}
f_{\rm NS} \simeq \frac{1}{\pi}\left(\frac{2GM}{a_{\rm f}^3}\right)^{1/2} \simeq 95\,{\rm Hz}\,\left(\frac{M}{0.1M_{\odot}}\right)\left(\frac{a_{\rm f}}{2R_{\rm NS}}\right)^{-3/2},
\end{equation}

close to the peak of the sensitivity band of ground-based GW detectors such as LIGO.  The maximum strain amplitude, as seen by an optimally oriented observer along the binary angular momentum axis, is approximately given by

\begin{equation}
h_{\rm NS} = \frac{4(G\mathcal{M}_{\rm NS})^{5/3}(\pi f_{\rm NS})^{2/3}}{c^{4}D} \approx 2 \times 10^{-25}\left(\frac{M}{0.1M_{\odot}}\right)^{7/3}\left(\frac{a_{\rm f}}{2R_{\rm NS}}\right)^{-1}\left(\frac{D}{100\,{\rm Mpc}}\right)^{-1}
\end{equation}

where $\mathcal{M}_{\rm NS} = M/2^{1/5}$ is the chirp mass and $D$ is the source distance.  
We have normalized the latter to a distance of 100 Mpc, roughly corresponding to the nearest collapsar event each year, 
given an assumed collapsar volumetric rate of $\approx 300$ Gpc$^{-3}$ yr$^{-1}$.  

The NS binary of mass $2M \ll M_{\bullet}$ orbits around the BH, emitting its own chirp GW signal of frequency

\begin{equation}
f_{\rm BH} \simeq \frac{1}{\pi}\left(\frac{GM_{\bullet}}{r^3}\right)^{1/2} \simeq 7.8\,{\rm Hz}\,\left(\frac{M_{\bullet}}{3M_{\odot}}\right)^{-1}\left(\frac{r}{200R_{\rm g}}\right)^{-3/2},
\end{equation}

with a corresponding maximum strain amplitude

\begin{equation}
h_{\rm BH} = \frac{4(G\mathcal{M}_\bullet)^{5/3}(\pi f_{\rm BH})^{2/3}}{c^{4}D} \approx 2\times 10^{-23}\left(\frac{M}{0.1M_{\odot}}\right)\left(\frac{M_{\bullet}}{3M_{\odot}}\right)^{-1/3}\left(\frac{r}{200R_{\rm g}}\right)^{-1}\left(\frac{D}{100\,{\rm Mpc}}\right)^{-1}.
\end{equation}

The merger with the BH will take place on a timescale,

\begin{equation}
\tau_{\rm BH} \simeq  \frac{5}{256}\frac{c^{5}r^{4}}{G^{3}M_{\bullet}^{2}(2M)} \approx 7\times 10^{3}\,{\rm s}\,\left(\frac{r}{200R_{\rm g}}\right)^{4}\left(\frac{M_{\bullet}}{3M_{\odot}}\right)^{2}\left(\frac{M}{0.1M_{\odot}}\right)^{-1}.
\end{equation}
For most initial separations $a_0$ satisfying $a_f \lesssim a_0 \lesssim R_{\rm H}$ we see that the NS binary will merge before the binary itself would migrate significantly towards the BH, i.e. $\tau_{\rm NS} < \tau_{\rm BH}$. Further, both events may occur after most of the collapsing star has already been accreted by the central BH or otherwise unbound in the explosion.

Combining results (for $a_{\rm f} = 2R_{\rm NS}$), we see that
\begin{equation}
\frac{h_{\rm NS}}{h_{\rm BH}} \approx 8\times 10^{-3}\left(\frac{M}{0.1M_{\odot}}\right)^{4/3}\left(\frac{r}{200R_{\rm g}}\right)\left(\frac{M_{\bullet}}{3M_{\odot}}\right)^{1/3},
\end{equation}
and
\begin{equation}
\frac{f_{\rm NS}}{f_{\rm BH}} \approx 12\left(\frac{M}{0.1M_{\odot}}\right)\left(\frac{M_{\bullet}}{3M_{\odot}}\right)\left(\frac{r}{200R_{\rm g}}\right)^{3/2}.
\end{equation}

Because typically $h_{\rm NS} \ll h_{\rm BH}$ but $f_{\rm NS} \gg f_{\rm BH}$, the NS binary waveform will appear as a low-amplitude high frequency 
``modulation'' on top of the slower and lower-frequency BH-NS merger. 

\begin{figure}[htbp]
\centering
\includegraphics[width=0.8\textwidth,clip=true]{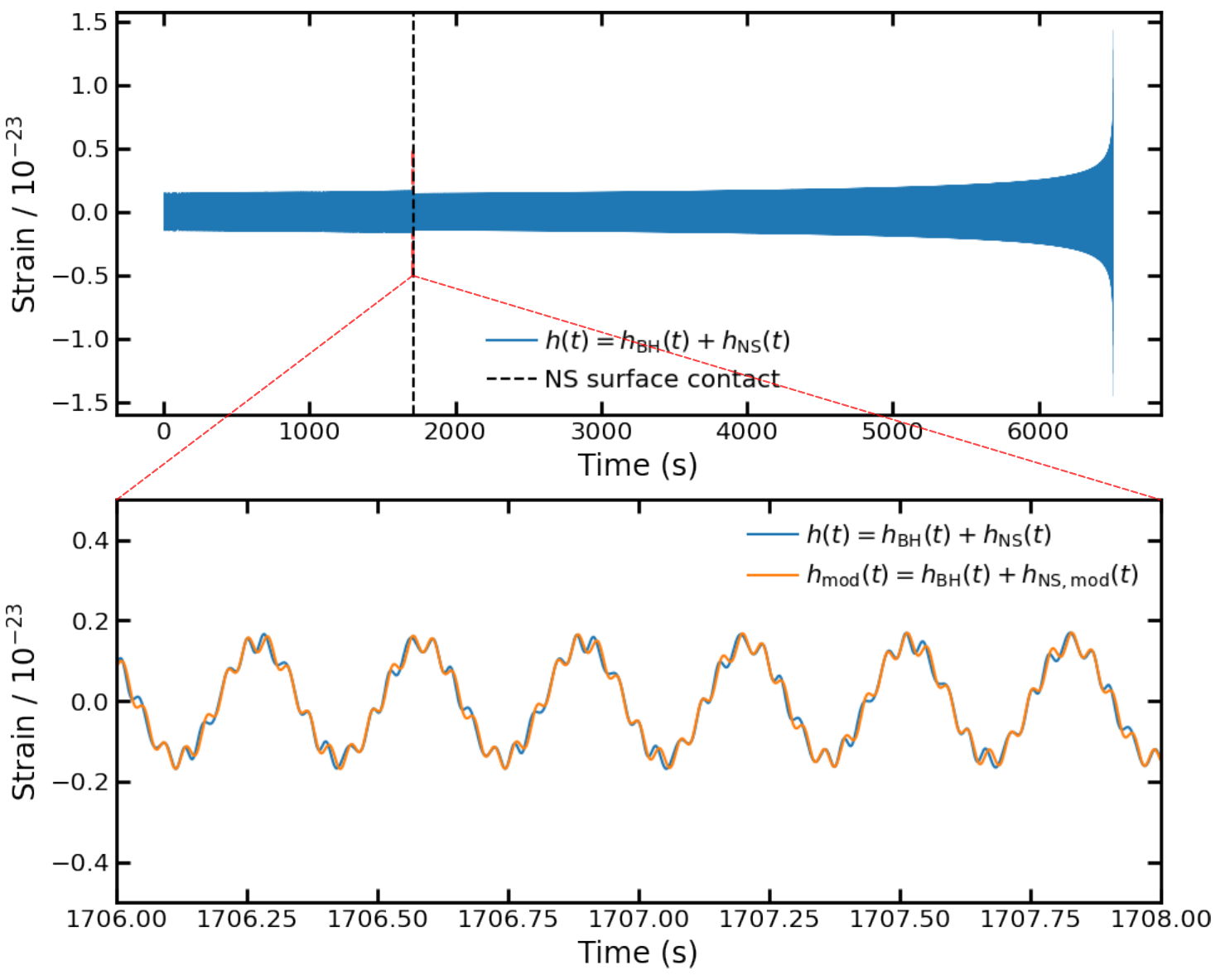}
\caption{Approximate gravitational waveform showing the combined strain $h_{\rm tot}(t)$ produced by the chirp $h_{\rm NS}(t)$ from a binary composed of two $0.1M_{\odot}$ NS in orbit around a $3M_{\odot}$ BH starting at $r = 200 R_{\rm g}$ and NSs at Hill radius separation, calculated by adding two point-mass waveforms in the 1st order post-Newtonian limit for an observer with an optimal orientation perpendicular to the angular momentum axes of both mergers. The top panel shows a zoom-out version starting from $t=0$ and ends at the final BH-NS coalescence, with a vertical dashed line showing the termination of the NS-NS merger.  The lower panel shows a zoomed in version of $h_{\rm tot}(t)$, 
as well as comparison with effects of including Doppler modulation of the NS-NS merger signal due to the binary's orbital motion (Eq.~\eqref{eq:hmod}). Our calculation neglects several other effects, including Doppler modulation of the strain amplitude and the Shapiro time delay.}
\label{fig:GWwaveform}
\end{figure}

As a quantitative example, 
we calculate the combined time-dependent strain evolution $h(t) = h_{\rm NS}(t) + h_{\rm BH}(t)$ for the two inspiralling binaries \citep{Peters64}, choosing a more realistic pre-factor of 2 that accounts for an appropriate average over orientation of the source and the antenna pattern of the detector \citep{1993ApJ...417L..17K},

\begin{eqnarray}
    h_{\rm BH}(t) = \frac{2(G\mathcal{M}_\bullet)^{5/3}(\pi f_{\rm BH}(t))^{2/3}}{c^{4}D}\cos \left(2\pi\int  f_{\rm BH}(t) dt\right)\\
    h_{\rm NS}(t) = \frac{2(G\mathcal{M}_{\rm NS})^{5/3}(\pi f_{\rm NS}(t))^{2/3}}{c^{4}D}\cos \left(2\pi\int  f_{\rm NS}(t) dt\right).
    \label{eqn:h(t)}
\end{eqnarray}

and the frequency grows approaching the merger as:
\begin{equation}
    f_{\rm BH}(t) = f_{\rm BH} \left(\dfrac{\tau_{\rm BH} - t}{\tau_{\rm BH}}\right)^{-3/8}, f_{\rm NS}(t) = f_{\rm NS} \left(\dfrac{\tau_{\rm NS} - t}{\tau_{\rm NS}}\right)^{-3/8}.
    \label{eqn:f(t)}
\end{equation}

We terminate the inspiral signals once the orbital separations become smaller than $2 R_{\rm NS}$ and $R_{\rm NS}$ respectively, corresponding roughly to the point of surface contact/merger. 

We plot the total waveform $h(t)$ in Figure \ref{fig:GWwaveform} for $M = 0.1M_{\odot}$, $M_{\bullet} = 3M_{\odot}$.  
Consistent with the above estimates, even just before the binary NS merger $t \lesssim \tau_{\rm NS}$ (lower panel) the total GW strain is still dominated by the binary-BH merger, even though the latter peaks much later. 
Nevertheless, zooming in on the waveform (bottom panel) we see that the different frequencies associated with $h_{\rm NS}$ and $h_{\rm BH}$ still enable differentiating the two signals.

The frequency of the binary NS merger chirp prior to merger is Doppler modulated by the orbital motion of the binary around the BH (\citetalias{metzger2024}), 
with an orbital period $2/f_{\rm BH} \sim 0.01-1$ s 
and a maximum amplitude 
$\delta f/f \approx v_{\rm BH}/c \approx 0.07(r/200R_{\rm g})^{-1/2}$ (maximum for a viewer close to the binary equatorial plane) where $v_{\rm BH} = (G \mathcal{M}_\bullet \pi f_{\rm BH})^{1/3}$ 
is the orbital velocity around the central BH, 
in effect leading to a modified binary NS strain which can approximately be written:

\begin{equation}
    h_{\rm NS, mod}(t) \simeq  \frac{2(G\mathcal{M})^{5/3}(\pi f_{\rm NS
    }(t))^{2/3}}{c^{4}D}\cos \left[2\pi \int f_{\rm NS}(t) \left(1 + \dfrac{v_{\rm BH}(t)}{2 c}\sin \left(\pi \int f_{\rm BH} (t) dt \right) \right) dt\right]
    \label{eq:hmod}
\end{equation}

We choose $\delta f/f  = v_{\rm BH}/2c $ for a representative inclination of $30^\circ$.  An orange solid line in the lower panel of Fig.~\ref{fig:GWwaveform} shows this waveform including this maximal Doppler modulation correction with roughly a period corresponding to twice that of the GW signal generated by the binary's inspiral towards the central BH. 

Optimistically, if template waveforms can be constructed and matched filtering techniques applied, one could attain an effective strain amplitude
\begin{equation}
    h_{c, \rm BH} =   \sqrt{\dfrac{f_{\rm BH}^2}{\dot{f}_{\rm BH}}} h_{\rm BH},\,\,\,\, h_{c, \rm NS} =  \sqrt{\dfrac{f_{\rm NS}^2}{\dot{f}_{\rm NS}}} h_{\rm N S},
    \label{eqn:hc(f)}
\end{equation}

enhanced roughly by the square root of the number of oscillation cycles spent in the observing band \citep{Flanagan&Hughes98}.
Figure \ref{fig:GW} shows that the predicted strain amplitudes could enable detection by Advanced LIGO \citep{Abbott_subsolar_22} with moderate to high signal-to-noise ratios of $\sim few-10$. 
Third-generation observatories like Einstein Telescope \citep{Branchesi+23} and Cosmic Explorer \citep{Evans+21} would improve this by roughly an order of magnitude.  

However, the complexities of the waveforms beyond those we have modeled could make it difficult to perform matched filtering across the entire parameter space of potential hierarchically merging systems.  Though instructive, the single equal-mass binary merger considered above is but one relatively simple example of what could be a much more complex GW waveform from a hierarchical chain of disk-embedded mergers, spanning an almost limitless parameter space of possibilities.  Our analysis above also does not account for several physical effects that will further complicate the gravitational waveform, such as gas-assisted orbital evolution, tidal interactions, and Shapiro time delays (particularly for edge on merger), among other effects.




\end{document}